\documentclass[aps,prd,reprint,twocolumn,superscriptaddress,preprintnumbers,nofootinbib]{revtex4-1}

\usepackage{amsthm}
\usepackage{amsmath}
\usepackage{graphicx}
\usepackage{slashed}
\usepackage{amssymb}
\usepackage{dsfont}
\usepackage{float}
\usepackage[colorlinks=True, citecolor=blue, urlcolor=blue, linkcolor=blue]{hyperref}
\usepackage[braket,qm]{qcircuit}
\usepackage{lineno}
\usepackage[dvipsnames]{xcolor}

\newcommand*\diff{\mathop{}\!\mathrm{d}}

\newcommand{\nn}{\nonumber}

\newcommand{\pp}[2]{\frac{\partial #1}{\partial #2}}
\newcommand{\be}{\begin{eqnarray}}
\newcommand{\ee}{\end{eqnarray}}
\newcommand{\ma}{\mathrm}
\newcommand{\ml}{\mathcal}

\newcommand{\Tr}{\mathrm{Tr}}

\newcommand{\Ncycle}{\ensuremath{N_{\rm{cycle}}}}
\DeclareMathOperator{\sign}{sign}

\begin{document}

\title{Quantum simulation of non-equilibrium dynamics and thermalization in the Schwinger model}

\author{Wibe A. de Jong}
\email{wadejong@lbl.gov}
\affiliation{Computational Research Division, Lawrence Berkeley National Laboratory, Berkeley, CA 94720, USA}

\author{Kyle Lee}
\email{kylelee@lbl.gov}
\affiliation{Nuclear Science Division, Lawrence Berkeley National Laboratory, Berkeley, California 94720, USA}
\affiliation{Physics Department, University of California, Berkeley, CA 94720, USA}

\author{James Mulligan}
\email{james.mulligan@berkeley.edu}
\affiliation{Nuclear Science Division, Lawrence Berkeley National Laboratory, Berkeley, California 94720, USA}
\affiliation{Physics Department, University of California, Berkeley, CA 94720, USA}

\author{Mateusz P\l osko\'n}
\email{mploskon@lbl.gov}
\affiliation{Nuclear Science Division, Lawrence Berkeley National Laboratory, Berkeley, California 94720, USA}

\author{Felix Ringer}
\email{fmringer@lbl.gov}
\affiliation{Nuclear Science Division, Lawrence Berkeley National Laboratory, Berkeley, California 94720, USA}

\author{Xiaojun Yao}
\email{xjyao@mit.edu}
\affiliation{Center for Theoretical Physics, Massachusetts Institute of Technology, Cambridge, MA 02139, USA}

\date{\today}
\preprint{MIT-CTP/5308}

\begin{abstract}
We present simulations of non-equilibrium dynamics of quantum field theories on digital quantum computers. As a representative example, we consider the Schwinger model, a 1+1 dimensional U(1) gauge theory, coupled through a Yukawa-type interaction to a thermal environment described by a scalar field theory. We use the Hamiltonian formulation of the Schwinger model discretized on a spatial lattice. With the thermal scalar fields traced out, the Schwinger model can be treated as an open quantum system and its real-time dynamics are governed by a Lindblad equation in the Markovian limit. The interaction with the environment ultimately drives the system to thermal equilibrium. In the quantum Brownian motion limit, the Lindblad equation is related to a field theoretical Caldeira-Leggett equation. By using the Stinespring dilation theorem with ancillary qubits, we perform studies of both the non-equilibrium dynamics and the preparation of a thermal state in the Schwinger model using IBM's simulator and quantum devices. The real-time dynamics of field theories as open quantum systems and the thermal state preparation studied here are relevant for a variety of applications in nuclear and particle physics, quantum information and cosmology.
\end{abstract}

\maketitle

{\it Introduction.} Quantum computing has emerged in recent years as a promising approach to solve a variety of classically intractable problems, due to considerable progress in hardware and algorithms~\cite{Devoret2013, annurev-conmatphys-031119-050605, doi:10.1063/1.5088164, google_supremacy}.
In particular, {\it quantum simulations} of real-time dynamics of systems where the classical computational cost scales exponentially with the system size may become tractable in the near or mid-term future~\cite{PhysRevB.101.184305, Smith2019}.
In high-energy and nuclear physics, a number of quantum computing applications have been proposed~\cite{Preskill_2018,Kaplan:2017ccd,Preskill:2018fag,Lamm:2018siq,Dumitrescu:2018njn,Roggero:2019myu,Bauer:2019qxa,Mueller:2019qqj,Wei:2019rqy,Holland:2019zju,Klco:2019evd,Avkhadiev:2019niu,DeJong:2020riy,Liu:2020eoa,Kreshchuk:2020dla,Davoudi:2020yln,Briceno:2020rar,Echevarria:2020wct,Chang:2020iwh,Hubisz:2020vhx,Cohen:2021imf,Barata:2021yri,Ramirez-Uribe:2021ubp,Li:2021kcs}. For applications in quantum field theories (QFTs), the Hamiltonian formulation of field theories~\cite{Kogut:1974ag} leads to exponentially large Hilbert spaces such that simulations may only become feasible with the advancement of quantum computing. In Refs.~\cite{Jordan:2011ci,Jordan:2011ne,Jordan:2014tma,Jordan:2017lea}, it was shown that scattering processes in scalar and purely fermionic field theories can be simulated efficiently with quantum computers, and they belong to the bounded-error quantum polynomial time (BQP) complete complexity class. Significant progress toward simulating field theories with quantum computing has been made over the past decade. Together with algorithmic advancements, first computations of quantum field theories for closed systems have been performed using real quantum devices~\cite{Horn:1981kk, Orland:1989st, Chandrasekharan:1996ih,Wiese:2013uua,Muschik:2016tws,Martinez:2016yna,Ercolessi:2017jbi,Klco:2018kyo,Raychowdhury:2018osk,Klco:2018zqz,Magnifico:2019kyj,Chakraborty:2020uhf,Shaw:2020udc,Kharzeev:2020kgc,Ikeda:2020agk,Alexandru:2019nsa,Barata:2020jtq,Harmalkar:2020mpd,Bauer:2021gup,Ciavarella:2021nmj}. See also Refs.~\cite{Banuls:2019bmf,Cloet:2019wre,Zhang:2020uqo} for recent reviews.

For most applications, it is crucial to prepare the initial state efficiently, which often is the ground state for real-time evolution in vacuum or a thermal state for real-time dynamics at finite temperature. Several approaches have been proposed to prepare a thermal state such as the quantum Metropolis algorithm~\cite{temme2011quantum}, the imaginary time evolution method~\cite{motta2020determining}, and the coupling with a heat bath~\cite{zalka1998simulating,terhal2000problem,wang2011quantum, PhysRevResearch.2.023214}. The last approach that is based on open quantum system formalism, or more generally, non-equilibrium dynamics of quantum systems, plays an important role in many physical systems.
Open quantum systems are relevant in high-energy and nuclear physics~\cite{Young:2010jq,Akamatsu:2011se,Gossiaux:2016htk,Brambilla:2017zei,Yao:2018nmy,Miura:2019ssi,Sharma:2019xum,Vaidya:2020cyi,Yao:2020xzw,Yao:2020eqy,Akamatsu:2020ypb,Brambilla:2020qwo,Yao:2021lus,Lehmann:2020fjt,Neill:2015nya,Armesto:2019mna,Li:2020bys}, cosmology~\cite{Boyanovsky:2015tba,Burgess:2015ajz,Shandera:2017qkg,Zarei:2021dpb,Cohen:2020php}, dark matter~\cite{Binder:2020efn}, and quantum information science~\cite{cleve_et_al:LIPIcs:2017:7477}. In particular, in ultra-relativistic heavy-ion collisions, probes of the quark-gluon plasma (QGP) such as heavy quark bound states or jets can be described as open quantum systems~\cite{Young:2010jq,Akamatsu:2011se,Gossiaux:2016htk,Brambilla:2017zei,Yao:2018nmy,Miura:2019ssi,Sharma:2019xum,Vaidya:2020cyi,Yao:2020xzw,Yao:2020eqy,Akamatsu:2020ypb,Brambilla:2020qwo,Yao:2021lus,Lehmann:2020fjt}. These studies are also closely related to the more general question on how the early stages of heavy-ion collisions 
form a QGP that is close to thermal equilibrium~\cite{Berges:2020fwq}.

\begin{figure}[t!]
\includegraphics[scale=0.25]{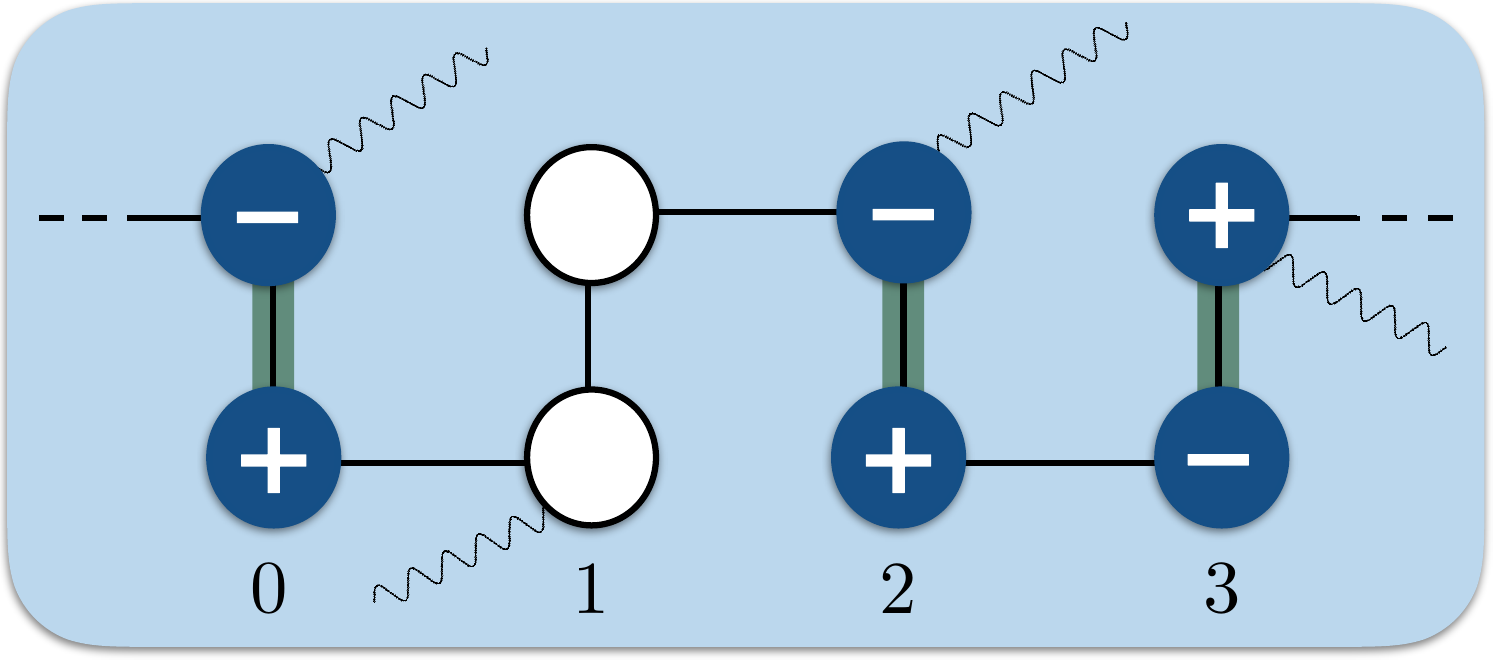}
\caption{Illustration of an example state in the Schwinger model with four spatial lattice sites (eight fermion sites). Empty circles indicate unoccupied sites while a circle with $-(+)$ represents a fermion site occupied by an electron (positron). The electric field is indicated by the green lines between occupied fermion sites. The wavy lines represent interactions with the environment. The dashed endpoints indicate periodic boundary conditions.~\label{fig:schwinger}}
\end{figure}

In this letter, we carry out quantum simulations of open systems described by QFTs for the first time. 
We consider a $1+1$ dimensional U($1$) gauge theory, the Schwinger model~\cite{Schwinger:1962tp,Coleman:1975pw}, as the open system coupled to a thermal environment consisting of scalar fields in $1+1$ dimensions. The Schwinger model serves as a compelling example for our studies
as it exhibits several features which are also present in quantum chromodynamics (QCD), such as confinement and spontaneous chiral symmetry breaking~\cite{Kharzeev:2012re,Loshaj:2014aia,Calzetta:2008iqa,Coleman:1976uz}. Due to its relative simplicity and important role in improving our understanding of more complex theories such as QCD, various studies have recently been carried out to investigate the real-time dynamics of the Schwinger model as a closed system~\cite{Muschik:2016tws,Martinez:2016yna,Klco:2018kyo,Chakraborty:2020uhf,Shaw:2020udc}. More recently, thermalization dynamics of the Schwinger model was studied on an analog quantum computer~\cite{Zhou:2021kdl}. In this work, we set up the relevant formalism to study field theoretical non-equilibrium dynamics. This formalism represents an important step toward studies of real-time dynamics of QCD in a thermal environment. In addition, we present results using the digital quantum devices accessible through the IBM Quantum (IBMQ) platform. Also, our study demonstrates a method for preparing an initial thermal state - an essential ingredient in applications of quantum computing to systems at finite temperature.

{\it Discretized Hamiltonian of the Schwinger model.} The Lagrangian of the (massive) Schwinger model is given by~\cite{Schwinger:1962tp,Coleman:1975pw}
\be\label{eq:Lagrangian}
\ml{L} = \overline{\psi} \big( i\slashed{D} - m \big) \psi - \frac{1}{4}F^{\mu\nu} F_{\mu\nu} \,,
\ee
where $\slashed{D} = \gamma^\mu D_\mu$ with $\{\gamma^\mu,\gamma^\nu\}=2g^{\mu\nu}$, the covariant derivative is $D_\mu = \partial_\mu - i e A_\mu$ and the field strength tensor is $F_{\mu\nu} = \partial_\mu A_\nu - \partial_\nu A_\mu$. The fermion field has two components $\psi=(\psi_u,\psi_d)^T$, where $u,d$ represent the upper and lower components, respectively. Furthermore, $m$ and $e$ denote the mass and the charge of the fermion, respectively.

To simulate the real-time dynamics of the Schwinger model on a quantum computer, we employ the Kogut-Susskind Hamiltonian formulation of Ref.~\cite{Kogut:1974ag} and discretize the field theory on a spatial lattice with $N$ sites. We employ periodic boundary conditions, which allows for the projection onto a reduced Hilbert space with definite momentum and parity. From the Lagrangian in Eq.~\eqref{eq:Lagrangian}, the discretized Hamiltonian can be obtained by choosing the axial gauge $A_0=0$, using staggered fermions~\cite{Casher:1973uf,Kogut:1974ag,Banks:1975gq}, and applying the Jordan-Wigner transformation~\cite{Jordan:1928wi}, which is reviewed in Appendices~\ref{app:a} and~\ref{app:b} in detail. The Hamiltonian can then be written as
\begin{align}\label{eq:H_S}
    H_S=
    &\,
    \frac{1}{2a} \sum_{n=0}^{N_{f}-1}\left(\sigma^{+}(n) L_{n}^{-} \sigma^{-}(n+1)+\sigma^{+}(n+1) L_{n}^{+} \sigma^{-}(n)\right)
    \nonumber\\&
    +\sum_{n=0}^{N_{f}-1}\left(\frac{ae^2}{2}\ell_{n}^{2}+m (-1)^{n} \frac{\sigma_{z}(n)+1}{2}\right) \,,
\end{align}
where $a$ is the lattice spacing, $n$ labels the fermion lattice sites, and $N_f=2N$ is the total number of fermion lattice sites.
See Fig.~\ref{fig:schwinger} for an example with $N_f=8$. The continuum theory is recovered in the limits $a\to 0$ and $N_f \to \infty$, such that $a N_f$ is fixed. Here $\sigma^{\pm}(n)=(\sigma_x(n) \pm i \sigma_y(n))/2$ and $\sigma_{x,y,z}(n)$ are the Pauli matrices at fermion site $n$. The operators $L^{\pm}_n$ are the raising/lowering operators for a quantum system with the eigenstates $|\ell_n\rangle$ associated with the eigenvalues $\ell_n$. The eigenvalues $\ell_n$ correspond to the electric flux between the fermion sites $n$ and $n+1$ while the ladder operators $L^{\pm}_n$ correspond to the gauge link between $n$ and $n+1$, which increases or decreases the electric flux by one unit. The subscript $S$ of the Hamiltonian in Eq.~(\ref{eq:H_S}) indicates that the Schwinger model will serve as the system interacting with a thermal environment, see Eq.~(\ref{eq:Hamiltonian_3}) below.

\begin{figure}[t!]
\includegraphics[scale=.26]{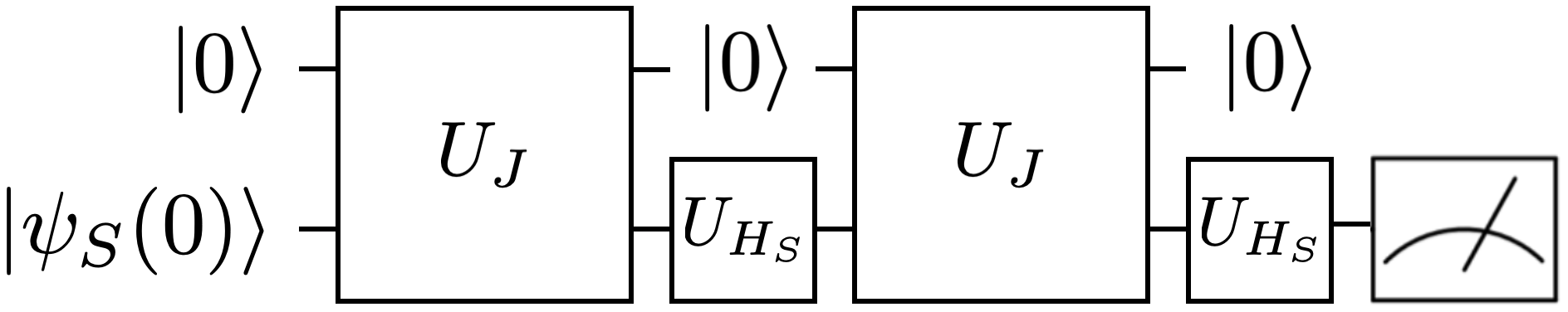}
\caption{Quantum algorithm to simulate the time evolution governed by the Lindblad equation using the Stinespring dilation theorem~\cite{nielsen_chuang_2010} for two cycles with time step $\Delta t$. The two unitary operators are given by $U_J=\exp{(-iJ\sqrt{\Delta t})}$ and $U_{H_S}=\exp{(-iH_S\Delta t)}$. The ancilla qubit is reset after each cycle. ~\label{fig:algorithm}}
\end{figure}

The lattice formulation of the Schwinger model varies in literature in how the infinite number of states of the gauge field are treated. Here we follow the setup of Ref.~\cite{Klco:2018kyo}, where a finite-dimensional representation of the gauge degrees of freedom is achieved by imposing a cutoff on the total electric flux. We find the following closed form for the number of physical states that satisfy Gauss's law with $|\ell_n|\leq 1$,
\begin{align}\label{eq:combinatorics}
\sum_{M=1}^{N} \frac{2N}{M}\sum_{K=0}^{N-M}\binom{M-1+K}{M-1}\binom{2N-2K-M-1}{M-1} + 3\,,
\end{align}
which is derived in Appendix~\ref{app:c}
In the following, we will focus on the Hilbert space projected onto positive-parity and zero-momentum states with $|\ell_n|\leq 1$ and $\sum_n|\ell_n|< N_f$. Constructions of these states and the matrix forms of the relevant Hamiltonians and measurement operators can be found in Appendix~\ref{app:b}. 

\begin{figure*}[!t]
\includegraphics[scale=0.5]{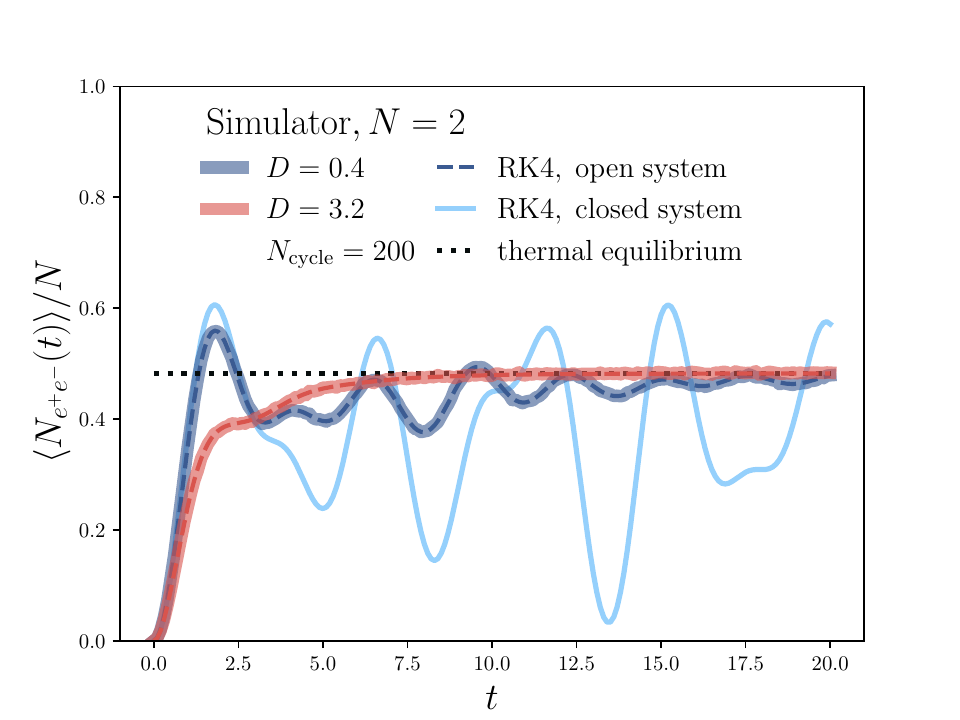}
\hspace*{.5cm}
\includegraphics[scale=0.5]{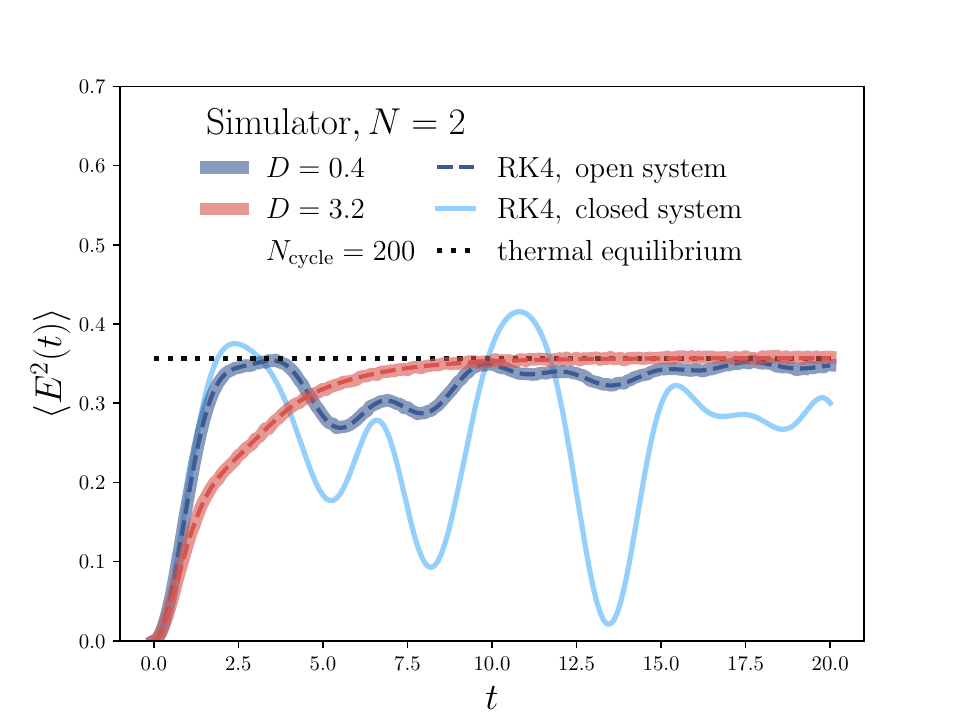}
\caption{Quantum simulation of non-equilibrium dynamics in the Schwinger model: $\langle N_{e^+e^-} \rangle$ (left) and $\langle E^2 \rangle$ (right) using the quantum circuit in Fig.~\ref{fig:algorithm} for two spatial lattice sites with $\Ncycle=200$, $\beta=0.1a$, $e=1/a$, $m=0.1/a$, $a=1$ and different system-environment couplings. The time $t$ is in units of $a$. For comparison, we also show a numerical solution (RK4) and the dotted line indicates the thermal equilibrium.~\label{fig:simulation}}
\end{figure*}

{\it Non-equilibrium dynamics in the quantum Brownian motion limit.}
We now consider the Schwinger model coupled to a thermal environment. The full Hamiltonian can be decomposed as
\begin{equation}\label{eq:Hamiltonian_3}
    H=H_S+H_E+H_I \,,
\end{equation}
where $H_S$ denotes the Hamiltonian of the system, i.e., the Schwinger model given in Eq.\ \eqref{eq:H_S}, $H_E$ is the environment Hamiltonian, and $H_I$ describes the interaction between the two. We take the environment to be a thermal scalar field theory and the coupling to the system to be a Yukawa-type interaction,
\be
H_E &=& \int \diff x \bigg[\frac{1}{2}\Pi^2 + \frac{1}{2}(\nabla \phi)^2 + \frac{1}{2}m_{\phi}^2\phi^2 + \frac{1}{4!} g \phi^4 \bigg]\,,\\
H_{I} &=& \lambda \int \mathrm{d} x\,  \phi(x) \overline{\psi}(x) \psi(x)
= \int \mathrm{d} x\, O_{E}(x) O_{S}(x)\,,\label{eq:H_I}
\ee
where $g>0$ and we define 
$O_E(x) = \lambda \phi(x)$ and $O_S(x) = \overline{\psi}(x) \psi(x)$.

We assume that the interaction $H_I$ is weak and that the environment is large enough that its change is negligible over the typical relaxation time of the system, i.e. we use the Markovian approximation. Then the full density matrix describing the system and the thermal environment factorizes $\rho(t) = \rho_S(t) \otimes \rho_E$, where $\rho_E = e^{-\beta H_E}/\Tr_E e^{-\beta H_E}$.
After tracing out the environmental degrees of freedom, the time evolution of the system density matrix $\rho_S={\rm Tr}_E[\rho]$ is governed by a Lindblad equation~\cite{KOSSAKOWSKI1972247,Lindblad:1975ef,Gorini:1976cm}.

Furthermore, we consider the quantum Brownian motion limit, which is valid when the environment correlation time $\tau_E$ is hierarchically smaller than both the relaxation time $\tau_R$ and the intrinsic time scale $\tau_S$ of the system~\cite{Yao:2021lus}. The condition $\tau_E\ll \tau_R$ is the Markovian condition mentioned above which is valid when $H_I$ is weak. The condition $\tau_E\ll \tau_S$ is equivalent to the hierarchy between the environment temperature $T$ and the characteristic energy gap of the system $\Delta E_S$: $T\gg \Delta E_S$. For QFTs, generally $\Delta E_S\to0$ in the continuum. The Schr\"odinger-picture Lindblad equation for the Schwinger model in the quantum Brownian motion limit can be written as 
\be
\label{eqn:lindblad_maintext}
\frac{\diff \rho_S(t)}{\diff t} = -i \big[H_S,\rho_S(t) \big] + L\rho_S(t) L^\dagger - \frac{1}{2}\big\{ L^\dagger L ,\rho_S(t) \big\}\,, \nn\\
\ee
which can be interpreted as a field theoretical Caldeira-Leggett equation~\cite{PhysRevLett.46.211} by dropping some of the higher order terms in the expansion of $\tau_E/\tau_S$. The corresponding Lindblad operator is given by
\be
L = \sqrt{a{N_f} D} \Big( O_S - \frac{1}{4T}\big[ H_S, O_S \big]
\Big)\,,
\ee
where $D$ is a function of $T$, $m_\phi$, $g$, and $\lambda$, given by
\be
D = \lambda^2 \int\diff t \diff x\, \Tr_E \big( \rho_E\, \phi(t,x) \phi(0,0) \big) \,.
\ee
The $D$ term is a two-point correlation function of the environment in momentum space. Here the frequency and momentum of the $D$ term are both zero. The frequency is zero because in the quantum Brownian motion limit, the energy gap is much smaller than the temperature, which allows an expansion in the ratio of the energy gap and the temperature. The momentum is also zero since here we focus on the Hilbert space consisting of only zero momentum states and thus there is no momentum transfer in dynamical processes.
In principle, one can calculate the environment correlation function, i.e., the $D$ term, by using thermal scalar field theory. Since it is independent of both frequency and momentum, we simply treat $D$ as an input parameter in the following numerical studies.
From Eq.~(\ref{eq:H_I}), we find $O_S=1/(2aN_f) \sum_n (-1)^n (\sigma_z(n)+1)  $.
Further details of the open quantum system formulation can be found in Appendix~\ref{app:d} and Ref.~\cite{Yao:2021lus}.

\begin{figure*}[!t]
\includegraphics[scale=0.5]{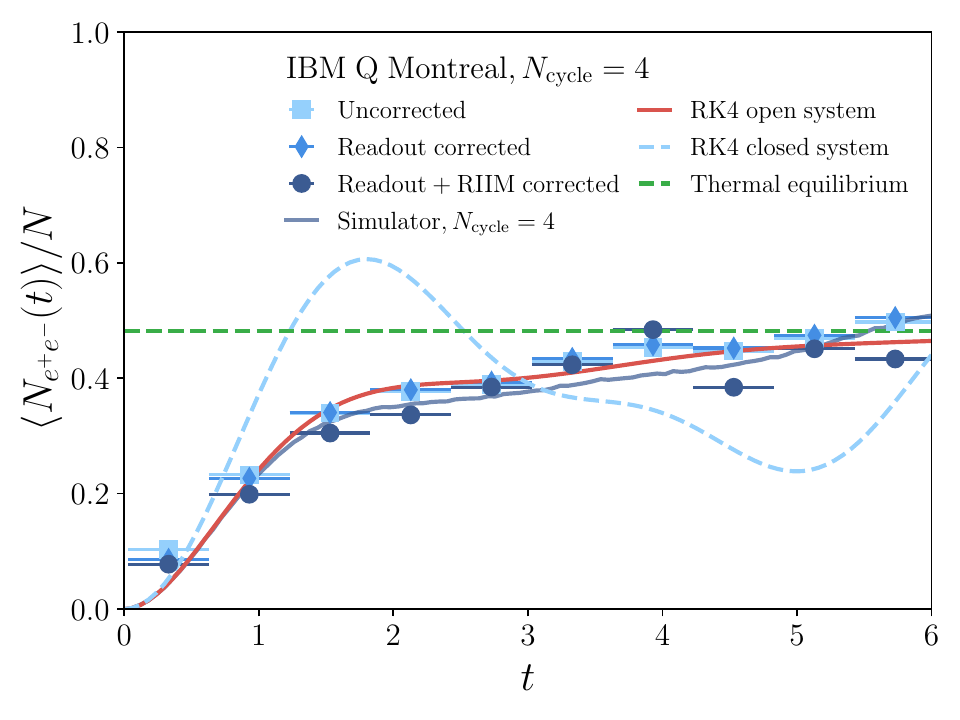}
\hspace*{.5cm}
\includegraphics[scale=0.5]{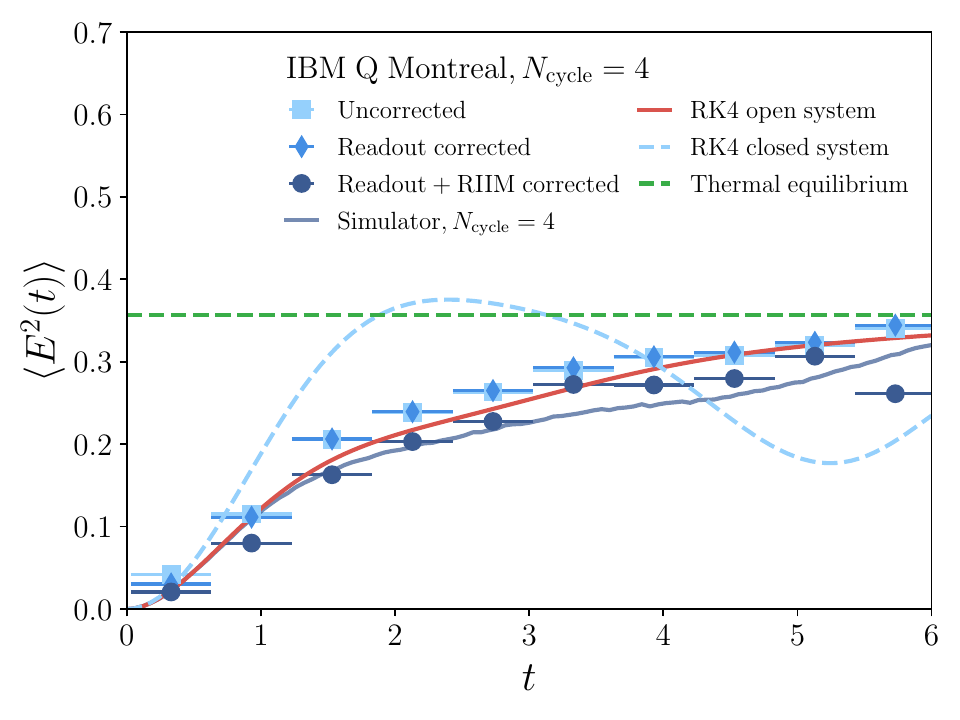}
\caption{Results from the \texttt{ibmq}$\bf{\_}$\texttt{montreal} device~\cite{IBMQMontreal} for $\langle N_{e^+e^-}\rangle$ (left) and $\langle E^2\rangle$ (right) for $N=2$ and $D=3.2$ with up to 4 cycles, see Fig.~\ref{fig:algorithm}. We include readout and CNOT error mitigation techniques as described in the text.~\label{fig:device}}
\end{figure*}

\begin{figure}[b]
\includegraphics[scale=0.53]{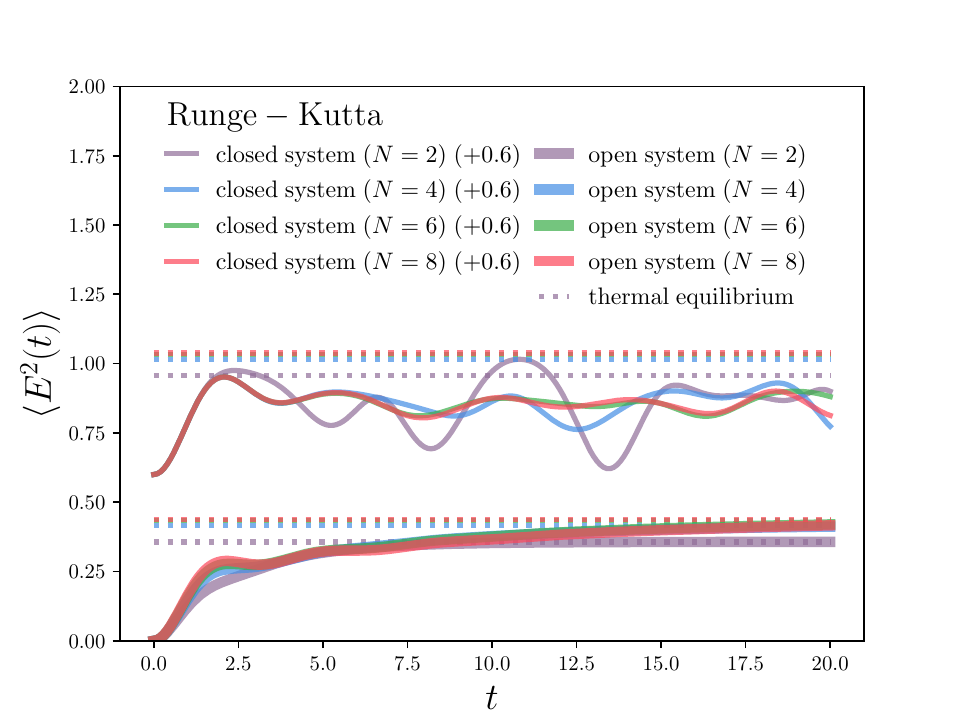}
\caption{Numerical solution of non-equilibrium dynamics in the Schwinger model: $\langle E^2 \rangle$ up to $N=8$ with $D=3.2$.
~\label{fig:simulation-N-dependence-E2}}
\end{figure}

{\it Quantum algorithm.} To simulate the non-unitary evolution in Eq.\ (\ref{eqn:lindblad_maintext}) of the system on a quantum computer, we apply the Stinespring dilation theorem~\cite{nielsen_chuang_2010,cleve_et_al:LIPIcs:2017:7477,DeJong:2020riy} to enlarge the Hilbert space, such that the system and additional ancillary qubits evolve unitarily together for small time steps. For the evolution from $0$ to $t$, we divide the length of the time interval into $N_{\ma{cycle}}$ time steps or ``cycles.'' For each cycle, we apply the algorithm with a time step $\Delta t= t/N_{\ma{cycle}}$ and the ancilla qubits are reset after each cycle. With qubit reset operations, we only need one ancilla qubit since Eq.~(\ref{eqn:lindblad_maintext}) has only one Lindblad operator. The quantum algorithm is shown schematically in Fig.~\ref{fig:algorithm} for two cycles. The initial state is given by $|\psi_S(0)\rangle \otimes |0\rangle_a$, where the initial state of the Schwinger model $|\psi_S(0)\rangle $ is chosen to be the unoccupied bare vacuum state while the ancilla is initialized in the $|0\rangle$ state. The $J$-operator is a $2\times2$ block matrix
\be
J = \begin{pmatrix}
0 & L^\dagger \\
L & 0 \\
\end{pmatrix}\,.
\ee
Other algorithms to simulate Lindblad equations are discussed in Refs.~\cite{cleve_et_al:LIPIcs:2017:7477,Hu:2019,PhysRevResearch.2.023214,headmarsden2020capturing,gupta2020optimal,PhysRevB.102.125112,ramusat2020quantum,metcalf2021quantum}.

{\it Simulation on IBMQ.} With the quantum algorithm discussed above, we begin by performing (noiseless) simulations using the IBMQ \texttt{qiskit} simulator~\cite{Qiskit}. We count all units in terms of $a$ and choose $e=1/a$, $m=0.1/a$ and $\beta\equiv 1/T=0.1a$. We then set $a=1$ and evolve in small time steps with $N_{\rm cycle}=200$. The results for two spatial lattice sites are shown in Fig.~\ref{fig:simulation}. We present results for both the expectation values of the number operators for $e^+e^-$ pairs, $\langle N_{e^+e^-} \rangle$ (left), and of the electric flux $\langle E^2 \rangle$ (right)
as a function of time for two different values of the correlator $D$. The result for the closed quantum system is shown for comparison. The open system starts to rapidly deviate from the closed system and eventually approaches the thermal equilibrium. Due to the interactions with the environment, the oscillations are damped. Both the oscillation damping rate and the system thermalization rate depend on the value of $D$. The results of the quantum algorithm are consistent with the results obtained with a $4$th order Runge-Kutta (RK4) method that solves Eq.~(\ref{eqn:lindblad_maintext}) classically. 
In Appendix~\ref{app:f}, we show simulation results of our quantum circuit up to $N=4$, which demonstrate similar agreement.

Next, we perform simulations using quantum devices from the IBMQ platform. We choose $N=2$ spatial lattice sites, which requires 2 qubits to represent the system and 1 additional qubit to simulate the interaction with the environment. We apply measurement error corrections using IBM's \texttt{qiskit-ignis} package~\cite{Qiskit}. In addition, we mitigate the Controlled NOT (CNOT) gate errors using the zero-noise extrapolation of Ref.~\cite{PhysRevA.102.012426}. To minimize error corrections, we opt for using a different ancilla qubit for every cycle instead of resetting a single ancilla qubit~\cite{rattew2021quantum}. We use the \texttt{qsearch} compiler of Ref.~\cite{2020arXiv201000215H} to efficiently map the two unitary operators $U_J$ and $U_{H_S}$, see Fig.~\ref{fig:algorithm}, to the basis gate set of IBMQ which consists of the single-qubit rotations RZ, SX, and X and the CNOT gate. One cycle (see Fig.~\ref{fig:algorithm}) consists of 7-13 CNOT gates and $\sim 100$ single-qubit gates. We show the results from the IBMQ device \texttt{ibmq}$\bf{\_}$\texttt{montreal}~\cite{IBMQMontreal} in Fig.~\ref{fig:device},
where we use a larger number of cycles (up to $N_{\rm{cycle}}=4$) as $t$ increases.
We find a very good agreement
between the quantum device and the noiseless circuit simulator.
The measurement error correction and the CNOT gate error mitigation
mildly improve the agreement, and generally have a small impact.
The performance deteriorates only slightly at later times due to the large number of CNOT gates. In addition, we observe that the quantum algorithm with 4 cycles gives a good approximation of the full result (labeled as ``RK4 open system'') up to $t\approx 6$. We are thus able to approximately prepare a thermal state of the Schwinger model from non-equilibrium dynamics. These results constitute the first studies of quantum simulations of quantum field theoretical non-equilibrium dynamics and thermalization.

In order for these simulations to describe physical systems, one needs to extrapolate
to the infinite volume and continuum limits.
As a first step, in Fig.~\ref{fig:simulation-N-dependence-E2}, we investigate finite volume effects of our results by simulating lattices with a different number of spatial sites $N$, for fixed lattice spacing $a$.
We plot the average electric field $\left< E^2 \right>$ using numerical methods (RK4) up to $N=8$,
and find the numerical solutions begin to converge as $N \rightarrow 8$. 
As indicated by the dashed horizontal lines, we find a mild dependence of the thermal equilibrium values on $N$.
Similar results for the number of electron-positron pairs are included in Appendix~\ref{app:g}.
In order to make the extrapolated results stable at higher temperatures, one needs to consider larger values of $N$ and include states with higher electric fluxes, since these states can then be excited more frequently.
For such high temperatures and large values of $N$, it will be eventually essential to use quantum computers to simulate the dynamics due to the exponential growth of the size of the physical Hilbert space.

{\it Conclusions.} We performed first quantum simulations of field-theoretical non-equilibrium dynamics of open quantum systems. We considered the Schwinger model discretized on a spatial lattice which interacts with a thermal scalar field theory. In the quantum Brownian motion limit, we derived the corresponding Lindblad evolution equation which can be cast in the form of a field-theoretical Caldeira-Leggett equation. We computed the non-unitary Lindblad evolution with IBM's simulator and with quantum hardware. We employed suitable optimization algorithms and error mitigation techniques and found good agreement with the Runge-Kutta solution which sets a benchmark for future studies.
In addition, we demonstrated a method for preparing thermal states for quantum computations of field theories -- a step that is important for studies of systems at finite temperature.
This work constitutes a starting point for simulations of the real-time evolution of non-equilibrium dynamics of quantum field theories with the ultimate goal of studying non-Abelian gauge theories in higher spatial dimensions.

\begin{acknowledgements}
{\it Acknowledgements.} We thank Mekena Metcalf, Krishna Rajagopal, Phiala Shanahan, and George Sterman for helpful discussions. We thank Marc Davis and Ethan Smith for help with the \texttt{qsearch} compiler~\cite{2020arXiv201000215H} and Alexander Barrett and Michael Earnest for helpful discussions on the combinatorics of the number of physical states of the Schwinger model. We acknowledge the use of IBM Quantum services for this work. The views expressed are those of the authors, and do not reflect the official policy or position of IBM or the IBM Quantum team. In this paper we used \texttt{ibmq}$\bf{\_}$\texttt{montreal}, which is one of the IBM Quantum Falcon Processors. This research used resources of the Oak Ridge Leadership Computing Facility, which is a DOE Office of Science User Facility supported under Contract DE-AC05-00OR22725. WDJ was supported by the U.S. Department of Energy, Office of Science, Office of Advanced Scientific Computing Research Accelerated Research in Quantum Computing program under contract DE-AC02-05CH11231. KL is supported by the US Department of Energy, Office of Nuclear Physics. JM, MP are supported by the U.S. Department of Energy, Office of Science, Office of Nuclear Physics, under the contract DE-AC02-05CH11231. FR is supported by LDRD funding from Berkeley Lab provided by the U.S. Department of Energy under Contract No. DE-AC02-05CH11231. XY is supported by the U.S. Department of Energy, Office of Science, Office of Nuclear Physics under grant Contract Number DE-SC0011090.
\end{acknowledgements}

\widetext
\appendix
\section{Hamiltonian formulation of the Schwinger model in the continuum}
\label{app:a}
The Lagrangian density of the (massive) Schwinger model is given by
\be\label{eq:Lagrangian1}
\ml{L} = \overline{\psi} \big( i\slashed{D} - m \big) \psi - \frac{1}{4}F^{\mu\nu} F_{\mu\nu} \,,
\ee
with $\slashed{D}=D_\mu\gamma^\mu$ and $\{\gamma^\mu, \gamma^\nu \} = 2g^{\mu\nu} $ for $\mu=0,1$. We use the metric that has $g^{00}=1$ and $g^{11}=-1$. The covariant derivative is $D_\mu = \partial_\mu - i e A_\mu$. The gamma matrices in 1+1-dimension can be chosen as
\be
\gamma^0=\sigma_z\,,\quad \gamma^1 = i\sigma_y\,,\quad \gamma^0\gamma^1 = \sigma_x \,,
\ee
where $\sigma_{x,y,z}$ are the Pauli matrices. The electromagnetic field strength tensor is $F_{\mu\nu} = \partial_\mu A_\nu - \partial_\nu A_\mu$. In 1+1 dimensions, the fermion field has two components: $\psi=(\psi_u,\psi_d)^T$. The mass dimensions of the fields and coupling constants are given by $[\psi]=[\overline{\psi}]=\frac{1}{2}$, $[A]=0$, $[e]=1$. The equation of motion associated with the gauge field $A_0$ is given by
\be
0=\frac{\partial\ml{L}}{\partial A_0} - \partial_1\frac{\partial\ml{L}}{\partial(\partial_1A_0)} = e\overline{\psi}\gamma^0\psi + \partial_1 F^{10} = e\psi^\dagger \psi + \partial_1 E \,,
\ee
which corresponds to Gauss's law and $E = F^{10}$ denotes the electric field.

For the Hamiltonian formulation of the Schwinger model, we will work in the axial gauge $A_0=0$. We denote the spatial component of the gauge field by $A_1=-A^1=A$ and we write the field strength tensor as $F^{10}=-F^{01}=E=\partial^0 A$. The canonical momenta $\Pi_\psi$ and $\Pi_A$ associated with the fermion and gauge fields, respectively, can be written as
\begin{align}
    \Pi_\psi & = \pp{\ml{L}}{(\partial^0 \psi)} = \overline{\psi} \, i \gamma^0 =  i\psi^\dagger \,,\\
    \Pi_A & = \pp{\ml{L}}{(\partial^0A^1)} = -E\,.
\end{align}
The nontrivial (anti-)commutation relations from canonical quantization are given by
\begin{align}
\{\psi(t,x), \Pi_\psi(t,y)\} & = i\{\psi(t,x), \psi^\dagger(t,y)\} =  i\delta(x-y) \,,\\
\quad [A^1(t,x), \Pi_A(t,y)] & = [A(t,x), E(t,y)] =  i\delta(x-y) \,.
\end{align}
The Hamiltonian density is then given by
\begin{equation}\label{eq:Hamiltonian_density}
    \ml{H} = \Pi_\psi \partial^0 \psi + \Pi_A \partial^0 A^1- \ml{L} \\
= -i\overline{\psi} \gamma^1 (\partial_1+ie A^1)\psi + m \overline{\psi}\psi + \frac12 E^2 \,.
\end{equation}
From now on, we will assume that all fields are in the Schr\"odinger picture. We note that in the Hamiltonian approach, Gauss's law is not generated by the equations of motion. Therefore, we have to impose $e\psi^\dagger\psi= - \partial_1 E$ when we construct the physical Hilbert space below.

For later convenience, we introduce the spatial Wilson line, the gauge link, as
\be
U(z,y) &=& \ml{P}\exp\bigg( -ie \int_y^z \diff x A^1(x) \bigg) = \ml{P}\exp\bigg( ie \int_y^z \diff x A_1(x) \bigg) \nn\\
\label{eqn:gauge_link}
&=& \sum_{n=0}^\infty \frac{(ie)^n}{n!} \int_y^z \diff x_1 \int_y^z \diff x_2 \cdots \int_y^z \diff x_n \ml{P}\big( A_1(x_1) A_1(x_2) \cdots A_1(x_n)  \big)\,,
\ee
where $\ml{P}$ denotes the path ordering operator. Using $[E(x), A_1(y)] = -i\delta(x-y)$ and
\be
\Big[ E(y), \ml{P}\big( A_1(x_1) A_1(x_2) \cdots A_1(x_n) \big) \Big]
&=& -i\delta(y-x_1) \ml{P}\big( A_1(x_2)A_1(x_3) \cdots A_1(x_n) \big) \nn\\
&&- i\delta(y-x_2) \ml{P}\big( A_1(x_1) A_1(x_3)A_1(x_4) \cdots A_1(x_n) \big) \nn\\
&& - \cdots \nn\\
&& - i\delta(y-x_n) \ml{P}\big( A_1(x_1) A_1(x_2) \cdots A_1(x_{n-1}) \big) \,,
\ee
we find
\be
\label{eqn:EU_commutation}
\big[ E(y), U(z,y) \big] = e U(z,y) \,.
\ee

\section{Lattice discretization}
\label{app:b}
We now review the lattice discretization of the Hamiltonian of the Schwinger model. We consider a 1-dimensional spatial lattice with lattice spacing $a$, while keeping time continuous. We label the lattice sites with an integer $n$ starting from $n=0$. The fields $\psi(x)$ and $A(x)$ at position $x=na$ are then labelled as $\psi(n)$ and $A(n)$ respectively. The discretized version of the fermionic part of the Hamiltonian $H_f=\int\diff x\,\ml{H}_f(x)$ can be written as
\be
\label{eqn:H_origin}
H_f &=& a\sum_n \Big( -i\overline{\psi}(n) \gamma^1 \frac{\psi(n+1) - \psi(n-1)}{2a} + m\overline{\psi}(n) \psi(n)
+e\overline{\psi}(n) \gamma^1 A^1(n) \frac{\psi(n+1)+\psi(n-1)}{2} 
\Big) + \ml{O}(a^3) \nn\\
&=& a\sum_n \Big( -i\psi^\dagger(n) \sigma_x \frac{\psi(n+1) - \psi(n-1)}{2a} + m\psi^\dagger(n) \sigma_z \psi(n)
-e\psi^\dagger(n) \sigma_x A(n) \frac{\psi(n+1)+\psi(n-1)}{2} 
\Big) + \ml{O}(a^3) \,,\qquad\ 
\ee
where we have used $A^1 = -A$.
To put the two-component fermion field $\psi=(\psi_u,\psi_d)^T$ on a lattice, we use the Kogut-Susskind staggered fermion approach~\cite{Casher:1973uf,Kogut:1974ag,Banks:1975gq}, where a field $\chi(n)$ with mass dimension $0$ is introduced as\footnote{An alternative definition of $\chi(n)$ is \begin{equation*}
\begin{cases}
\frac{\chi(2n)}{\sqrt{a}} = \psi_u(n)\,,  \\
\frac{\chi(2n+1)}{\sqrt{a}} = \psi_d(n)\,.
\end{cases}
\end{equation*}
In either definition of $\chi(n)$, the number of fermion lattice sites required to represent $N_f/2$ spatial sites is $N_f$.}
\be
\chi(n) = \sqrt{a} \, (\sigma_x)^n \psi(n) \,,\qquad\qquad
\chi^\dagger(n) = \sqrt{a} \, \psi^\dagger(n) (\sigma_x)^n \,.
\ee
The Hamiltonian can then be written as
\be
H_f &=& \frac{1}{2a}\sum_n\Big( 
- i\chi^\dagger(n) \chi(n+1)  -ae \chi^\dagger(n) A(n) \chi(n+1) + i \chi^\dagger(n) \chi(n-1) - ae \chi^\dagger(n) A(n) \chi(n-1) \nn\\
&& +2ma(-1)^n \chi^\dagger(n) \sigma_z \chi(n) + \ml{O}(a^2)
\Big) \,,
\ee
where we have used $\sigma_x^2=1$.
Using the definition of the gauge link in Eq.~(\ref{eqn:gauge_link}), we can write
\be
\chi^\dagger(n) U(n,n+1)\chi(n+1) &=& \chi^\dagger(n) \ml{P} e^{ie\int_{(n+1)a}^{na}\diff x A(x)} \chi(n+1) \nn\\
&=& \chi^\dagger(n) \chi(n+1) - iae \chi^\dagger(n) A(n) \chi(n+1) + \ml{O}(a^2) \,,
\ee
and similarly
\be
\chi^\dagger(n) U(n,n-1)\chi(n-1) = \chi^\dagger(n) \chi(n-1) + iae \chi^\dagger(n) A(n) \chi(n-1) + \ml{O}(a^2) \,.
\ee
Then we can write the Hamiltonian as
\be
\label{eqn:app_Hf_chi2}
H_f &=& \frac{1}{2a}\sum_n \Big(
-i\chi^\dagger(n) U(n,n+1) \chi(n+1) + i\chi^\dagger(n) U(n,n-1) \chi(n-1) + 2ma(-1)^n \chi^\dagger(n) \sigma_z \chi(n)
\Big) \nn\\
&=& \frac{1}{2a}\sum_n \Big(
- i\chi^\dagger_u(n) U(n,n+1) \chi_u(n+1) + i\chi^\dagger_u(n) U(n,n-1) \chi_u(n-1) + 2ma(-1)^n \chi^\dagger_u(n) \chi_u(n) \nn\\
&& \quad\quad\quad\ \ \,  -\, i\chi^\dagger_d(n) U(n,n+1) \chi_d(n+1) + i\chi^\dagger_d(n) U(n,n-1) \chi_d(n-1) - 2ma(-1)^n \chi^\dagger_d(n) \chi_d(n)
\Big) \,.
\ee
where in the last two lines we have explicitly written out the upper and lower components of the fermion field $\chi$. We note that the lower component $\psi_d$ at odd (even) sites behaves in the same way as the upper component $\psi_u$ at even (odd) sites. Therefore, we can discard the lower component $\psi_d$ and treat the field $\chi$ in the first line of Eq.~\eqref{eqn:app_Hf_chi2} as a single-component, Grassmann-valued field. The summation here is over $N_f$ fermion sites with spacing $a$ and two fermion sites per spatial lattice site as illustrated in Fig.~\ref{fig:schwinger}.

Next, we apply the Jordan-Wigner transformation~\cite{Jordan:1928wi} which maps the staggered fermion fields to spin matrices as
\be
\chi_u(n) \to  \bigg(\prod_{m < n} -i\sigma_z(m) \bigg) \sigma^-(n) \,,\quad\qquad
\chi^\dagger_u(n) \to \sigma^+(n) \bigg(\prod_{m < n} +i\sigma_z(m) \bigg)  \,,
\ee
where we define $\sigma^\pm(n) = (\sigma_x(n) \pm i \sigma_y(n) )/2$ as the ladder matrices at site $n$. 
To complete the lattice formulation of the Schwinger model, we also need to discretize the electric field and the gauge link. There are different approaches in the literature to treat the infinite dimensional Hilbert space of the gauge field. The so-called quantum link model~\cite{Horn:1981kk, Orland:1989st, Chandrasekharan:1996ih, Wiese:2013uua} replaces U($1$) gauge-fields by spin variables, which allows for finite, but non-unitary representations of the canonical commutation relations. In Refs.~\cite{Muschik:2016tws,Martinez:2016yna}, the gauge field is eliminated using Gauss's law at the cost of long-range interactions. In Refs.~\cite{Magnifico:2019kyj,Ercolessi:2017jbi}, the U($1$) gauge degrees of freedom are described by the finite dimensional implementation through the discrete group $\mathds{Z}_m$, where U($1$) is restored in the large-$m$ limit. In our work, we follow the approach of Ref.~\cite{Klco:2018kyo}. The Hilbert space is restricted to physical states which satisfy Gauss's law and an upper cutoff on the Hilbert space of the gauge field is imposed. The discretized version of the commutation relation in Eq.~(\ref{eqn:EU_commutation}) can be written as
\be
[E(n),U(n+1,n)] = e U(n+1,n) \,.
\ee
This quantum system for each $n$ can be solved like a harmonic oscillator:
\be
    E(n)|\ell_n\rangle &=& e \ell_n |\ell_n \rangle \,,\\
    U(n\pm 1,n)|\ell_n \rangle &=& |\ell_n \pm 1 \rangle \,.
\ee
Here $\ell_n=0,\pm 1,\pm 2,\cdots$ denotes the eigenvalue (up to the factor $e$) of the electric operator $E(n)$ at site $n$ and $|\ell_n\rangle$ denotes the corresponding eigenstate. Using the eigenstates as a basis, the electric field and the gauge link can be represented as
\be
E(n) \to e \ell_n\,,\quad U(n,n-1)\to L^+_{n-1}\,,\quad U(n,n+1)\to L^-_n
\ee
The $L_n^{\pm}$ operators raise/lower the electric flux on the link between the fermion sites $n$ and $n+1$ and act as $L_n^\pm|\ell_n\rangle=|\ell_n\pm 1\rangle$. Putting everything together, we finally obtain the discretized Hamiltonian of the Schwinger model
\be
H_S=
\frac{1}{2a} \sum_{n}\Big(\sigma^{+}(n) L_{n}^{-} \sigma^{-}(n+1)+\sigma^{+}(n) L_{n-1}^{+} \sigma^{-}(n-1)\Big)  + \frac{m}{2} \sum_n (-1)^{n} \big( \sigma_{z}(n)+1 \big) + \frac{ae^2}{2} \sum_{n} \ell_{n}^{2} \,.
\ee
We employ periodic boundary conditions such that the 1-dimensional chain shown in Fig.~\ref{fig:schwinger} effectively forms a circle. Note that in the main text, we redefine $n$ by $n+1$ for the second term in the Hamiltonian. As discussed earlier, Gauss's law $\partial_1 E = - e \psi^\dagger \psi$ has to be imposed to form physical states, which has the discrete form:
\be
E(n+1) - E(n) = - e \sigma^+(n) \sigma^-(n) - e \frac{(-1)^n - 1}{2} \,,
\ee
where the constant term appears in the staggered fermion approach.
Imposing Gauss's law significantly reduces the size of the Hilbert space, though the size of the Hilbert space of physical states still grows exponentially with the number of lattice sites. In fact, we find an analytical expression for the number of physical states, which will be discussed further below. 

For our numerical calculations, we project onto states with zero momentum $\mathbf{k}=0$ and positive parity. The zero-momentum states can be constructed by first defining equivalent classes under cyclic permutations. In each equivalent class, the states are related to each other via cyclic permutations. Then the symmetrized linear combination of all states in the same equivalent class gives one zero-momentum state. The parity transformation is defined by a reflection with respect to a given site. If a zero-momentum state is invariant under the parity transformation, it is a positive parity state itself. If a zero-momentum state becomes another zero-momentum state under the parity transformation, then their symmetrized linear combination gives a positive parity state. We have set up a Python code that can generate all the physical states, project onto the states onto zero-momentum and positive parity, and which computes the corresponding Hamiltonian matrix in the basis of these states. We verified the result of our code by an explicit calculation for $N=2,4$ spatial lattice sites.

With a cutoff $|\ell_n|\leq 1$ on the quantized electric flux, the Hamiltonian for two spatial lattice sites can be written as~\cite{Klco:2018kyo}
\begin{equation}\label{eq:Hamiltonianmatrix}
H_S^{\mathbf{k}=\mathbf{0},+}=\left(\begin{array}{ccccc}
-2m & \frac{1}{a} & 0 & 0 & 0 \\
\frac{1}{a} & \frac{ae^2}{2} & \frac{1}{\sqrt{2}a} & 0 & 0 \\
0 & \frac{1}{\sqrt{2}a} & ae^2+2m & \frac{1}{\sqrt{2}a} & 0 \\
0 & 0 & \frac{1}{\sqrt{2}a} & \frac{3ae^2}{2} & \frac{1}{\sqrt{2}a} \\
0 & 0 & 0 & \frac{1}{\sqrt{2}a} & 2ae^2-2m
\end{array}\right)  
\end{equation}
For four spatial lattice sites we have
\begin{align}\label{eq:Hamiltonianmatrix4}
&H_S^{\mathbf{k}=\mathbf{0},+}=
\nonumber\\&
\hspace{-0.5cm}\left(\begin{smallmatrix}
-4 m  & \frac{\sqrt{2}}{a} & 0    & 0  & 0  & 0  & 0  & 0  & 0  & 0  &  0 &  0 &  0 &  0 &  0 &  0 &0 &  0 &0 \\ 
 \frac{\sqrt{2}}{a} & \frac{ae^2}{2}-2 m & \frac{1}{a}   & \frac{1}{\sqrt{2}a} & \frac{1}{\sqrt{2}a} & \frac{1}{\sqrt{2}a} & 0  & 0  & 0  & 0  &  0 &  0 &  0 &  0 &  0 &  0 &0 &  0 &0 \\ 
 0    & \frac{1}{a}   & ae^2   & 0  & 0  & 0  & \frac{1}{2a} & \frac{1}{a}   & \frac{1}{2a}   & 0  &  0 &  0 &  0 &  0 &  0 &  0 &0 &  0 &0 \\ 
 0    & \frac{1}{\sqrt{2}a}   & 0    & ae^2 & 0  & 0   & 0 & 0  & \frac{1}{\sqrt{2}a}  & 0  &  0 &  0 &  0 &  0 &  0 &  0 &0 &  0 &0 \\ 
 0    & \frac{1}{\sqrt{2}a}   & 0    & 0  & ae^2 & 0  & 0   & \frac{1}{\sqrt{2}a}   &0 & 0  &  0 &  0 &  0 &  0 &  0 &  0 &0 &  0 &0 \\ 
 0    & \frac{1}{\sqrt{2}a}   & 0    & 0  & 0  & ae^2 & 0  & 0  & \frac{1}{\sqrt{2}a} & 0  &  0 &  0 &  0 &  0 &  0 &  0 &0 &  0 &0 \\ 
  0    & 0    & \frac{1}{2a} & 0  & 0  & 0  & \frac{3}{2} ae^2-2 m   & 0& 0  & \frac{1}{2a}  & \frac{1}{2a}   &  \frac{1}{2a}  &  0  &  0 &  0 &  0 &0 &  0 &0 \\ 
 0    & 0    & \frac{1}{a}   & 0 & \frac{1}{\sqrt{2}a}  & 0  & 0 & \frac{3}{2} ae^2+2 m  & 0  & \frac{1}{2a} &  0  &  \frac{1}{2a}  &  \frac{1}{a} & 0 & 0 &  0 &0 &  0 &0 \\ 
 0    & 0    & \frac{1}{2a} & \frac{1}{\sqrt{2}a}  & 0 & \frac{1}{\sqrt{2}a} & 0  & 0  & \frac{3}{2} ae^2+2 m & 0  &  \frac{1}{2a} &  0 &  0  &  0 &  0 &  0 &0 &  0 &0 \\ 
 0    & 0    & 0    & 0  & 0  & 0  & \frac{1}{2a}   & \frac{1}{2a}   & 0  & 2 ae^2  & 0 &  0 &  0 &   \frac{1}{2a} &   \frac{1}{2a} &  0 &0 &  0 &0 \\ 
 0    & 0    & 0    & 0  & 0  & 0  & \frac{1}{2a}   & 0   & \frac{1}{2a}  & 0  &   2 ae^2  &0&  0 &   0 &  0 &  0 &0 &  0 &0 \\ 
 0    & 0    & 0    & 0  & 0  & 0  & \frac{1}{2a}  & \frac{1}{2a}   & 0   & 0  &  0 &  2 ae^2 & 0 & \frac{1}{2a}  &  \frac{1}{2a} &  0 &0 &  0 &0 \\ 
  0    & 0    & 0    & 0  & 0  & 0  & 0 & \frac{1}{a}  & 0  & 0 &  0 &  0 &  2 ae^2+4 m & 0 &  \frac{1}{a} &  0 &0 &  0 &0 \\ 
0    & 0    & 0    & 0  & 0  & 0  & 0  & 0  & 0   &  \frac{1}{2a}  &  0  &  \frac{1}{2a}  &  0 &  \frac{5}{2} ae^2-2 m & 0 &  \frac{1}{2a} &  &  0 &0 \\ 
 0    & 0    & 0    & 0  & 0  & 0  & 0  & 0  & 0  & \frac{1}{2a} &   0 &  \frac{1}{2a}  &  \frac{1}{a} & 0 &  \frac{5}{2} ae^2+2 m & \frac{1}{a} &   \frac{1}{\sqrt{2}a} &  0 &0 \\ 
 0    & 0    & 0    & 0  & 0  & 0  & 0  & 0  & 0   & 0  &  0 &  0 &  0 &\frac{1}{2a}  &  \frac{1}{a}  &   3 ae^2 &0 & \frac{1}{a} &0 \\ 
 0    & 0    & 0    & 0  & 0  & 0  & 0  & 0  & 0  & 0  &  0 &  0 &  0 & 0 &   \frac{1}{\sqrt{2}a} &  0 & 3 ae^2 & \frac{1}{\sqrt{2}a} &0 \\ 
 0    & 0    & 0    & 0  & 0  & 0  & 0  & 0  & 0  & 0  &  0 &  0 &  0 &  0 &  0 & \frac{1}{a} &   \frac{1}{\sqrt{2}a} & \frac{7}{2} ae^2-2 m &   \frac{1}{a} \\ 
 0    & 0    & 0    & 0  & 0  & 0  & 0  & 0  & 0  & 0  &  0 &  0 &  0 &  0 &  0 &  0 &   0  & \frac{1}{a} & 4 ae^2-4 m  \end{smallmatrix}\right)  
\end{align}
Here the states are arranged in ascending order in terms of gauge fields and $e^+e^-$ pairs. We note that we find full agreement when comparing the ground state energy eigenvalues to the results given in Ref.~\cite{Klco:2018kyo}. For completeness, we also give the measurement operators $\hat A$ for the electric field and the number of $e^+e^-$ pairs in this basis, which are defined by
\be
\hat{A}_{E^2} &=& \frac{1}{2Na}\int \diff x\,E^2(x) = \frac{e^2}{2N}\sum_n \ell_n^2 \,,\\
\hat{A}_{N_{e^+e^-}} &=& \sum_{n,\,{\rm even}} \sigma^+(n) \sigma^-(n) \,,
\ee
respectively, where $2Na = a N_f $ is the total length of the spatial lattice. Here even lattice sites correspond to electrons.
For two spatial lattice sites we find
\be
    \hat A^{\mathbf{k}=\mathbf{0},+}_{E^2} &=& \frac{e^2}{4} {\rm diag}(0,1,2,3,4)\,,\\
    \hat A^{\mathbf{k}=\mathbf{0},+}_{N_{e^+e^-}} &=& {\rm diag}(0,1,2,1,0)\,,
\ee
and for four spatial lattice sites we have
\be
    \hat A^{\mathbf{k}=\mathbf{0},+}_{E^2} &=& \frac{e^2}{8} {\rm diag}(0,1,2,2,2,2,3,3,3,4,4,4,4,5,5,6,6,7,8)\,,\\
    \hat A^{\mathbf{k}=\mathbf{0},+}_{N_{e^+e^-}} &=& {\rm diag}(0,1,2,2,2,2,1,3,3,2,2,2,4,1,3,2,2,1,0)\,.
\ee
The observables that we study in the main text are defined by
\be
\langle E^2(t) \rangle &\equiv& \Tr(\rho_S(t) \hat{A}_{E^2}) \,, \\
\langle N_{e^+e^-}(t) \rangle &\equiv& \Tr(\rho_S(t) \hat{A}_{N_{e^+e^-}}) \,.
\ee

\section{Combinatorics of the Schwinger model}
\label{app:c}
In this section, we derive the combinatorial formula in Eq.~(\ref{eq:combinatorics}) which counts the number of physical states satisfying Gauss's Law for a given lattice size. We take the size of the spatial lattice to be $N$, which means there are $N$ electron and $N$ positron sites. We impose a cutoff on the electric flux at each link, i.e. $|\ell_n|\leq 1$ for $n = 0, 1,...,2N-1$. Then the number of physical states with $M$ pairs of $e^+e^-$, denoted by $D_{N,M}$, is given by, up to a symmetry factor that will be described below, the number of unique solutions $(x_1,x_2,...,x_M;y_1,y_2,...,y_M)$ of the partition equation
\be
\label{eq:Me+e-}
x_1+x_2+\cdots+x_M+y_1+y_2+\cdots+y_M = 2N \,,
\ee
where $x_i \in \{1,3,5,...\}$ and $y_i \in \{1,2,3,4,...\}$. Here, $x_i$ represents the distance between the $i^{\rm th}$ $e^+e^-$ pair connected by an electric field, which is always an odd integer. Moreover, $y_i$ represents the size of the $i^{\rm th}$ gap between different fermions where the electric flux is zero. The length of these gaps is always a positive integer. Using the ``stars-and-bars'' method in combinatorics~\cite{Feller2} to solve Eq.~\eqref{eq:Me+e-} and including the symmetry factor $2N/M$, we arrive at
\be
D_{N,M}=\frac{2N}{M}\sum_{K=0}^{N-M}\binom{M-1+K}{M-1}\binom{2N-2K-M-1}{M-1}, \qquad\text{for } M = {1,2,...,N}\,.
\ee
The $2N$ of the overall symmetry factor $2N/M$ can be understood as the number of cyclic permutations that creates unique configurations. The $1/M$ is a factor to correct for overcounting since $(x_1,..,x_M,y_1,...,y_M) = (x_2,...,x_M,x_1;y_2,...,y_M,y_1)=\cdots=(x_M,x_1,..,x_{M-1};y_M,y_1,...,y_{M-1})$ describe the same physical configuration. Finally, we arrive at the expression that gives the total number of physical configurations with the restriction $|\ell_n|\leq 1$:
\begin{align}
D_{N}\equiv& \sum_{M=1}^{N} D_{N,M} + D_{N,0} \nonumber\\
=& \sum_{M=1}^{N} \frac{2N}{M}\sum_{K=0}^{N-M}\binom{M-1+K}{M-1}\binom{2N-2K-M-1}{M-1} + 3\,.
\end{align}
We note that $D_{N,0} = 3$ for all $N$ which counts the states with no $e^+e^-$ pairs and the electric flux is everywhere $-1$ or $0$ or $1$.

With this closed formula, we give results of $D_N$ explicitly for different values of $N$ in table~\ref{tab:scaling}. We find agreement with Ref.~\cite{Klco:2018kyo} where the results up to $N=12$ were given.
\begin{table}[!th]
\centering
\begin{tabular}{|c|c|}
\hline
\ \  $N$\ \  & \ \ ${D}_N$ \\
 \hline
 1 & 5 \\
 2 & 13 \\
 4 & 117 \\
 6 & 1,186 \\
 8 & 12,389\\
 10 & 130,338\\
 12 & 1,373,466 \\ 
 14 & 14,478,659 \\ 
16 & 152,642,789 \\ 
18 & 1,609,284,589 \\ 
20 & 16,966,465,802 \\
\vdots &\vdots\\
50 & 37,495,403,206,807,318,414,369,013\\
\vdots &\vdots\\
100 & 1,405,905,261,641,056,248,331,375,526,910,312,847,554,957,270,229,877\\
 \hline
\end{tabular}
\caption{Number of physical states in the Hilbert space of the Schwinger model $D_N$ for $|\ell_n|\leq 1$ up to $N=100$ spatial lattice sites.~\label{tab:scaling}}
\end{table}

\section{Open quantum systems, Lindblad evolution and quantum Brownian motion}
\label{app:d}
We consider the Schwinger model coupled to a thermal scalar field. The Hamiltonian of the whole system, which consists of the system (the Schwinger model) and the environment (the thermal scalar field), can be written as
\be
H = H_S + H_E + H_I\,.
\ee
Here $H_S$ and $H_E$ denote the Hamiltonians of the system and the environment respectively. The Hamiltonian of the $1+1$ dimensional scalar field can be written as
\be
H_E = \int \diff x \bigg[\frac{1}{2}\Pi^2 + \frac{1}{2}(\nabla \phi)^2 + \frac{1}{2}m_{\phi}^2\phi^2 + \frac{1}{4!} g \phi^4 \bigg]\,,
\ee
where $\Pi$ is the canonical momentum conjugate to $\phi$ and $g>0$.
The interaction Hamiltonian is assumed to be a Yukawa-type coupling:
\begin{align}
    H_{I}=\lambda \int \mathrm{d} x\,  \phi(x) \overline{\psi}(x) \psi(x)=\int \mathrm{d} x\, O_{E}(x) O_{S}(x)\,,
\end{align}
where $O_E(x) = \lambda \phi(x)$ and $O_S(x) = \overline{\psi}(x) \psi(x)$. The time evolution of the whole system is given by the von Neumann equation
\be
\frac{\diff \rho(t)}{\diff t} = -i \big[ H, \rho(t) \big] \,.
\ee
Tracing out the environment degrees of freedom, we obtain the reduced evolution equation of the system (the Schwinger model) in the interaction picture
\be
\label{eqn:reduced}
\rho_S^{(\text{int})}(t) &=& \Tr_E \big( \rho^{(\text{int})}(t) \big) = \Tr_E \big( U(t) \rho^{(\text{int})}(0) U^{\dagger}(t) \big)\,, \\
U(t) &=& \ml{T} \exp\Big(-i \int_0^t H_I^{(\text{int})}(t') \diff t'\Big) \,.
\ee
Here $\ml{T}$ denotes the time ordering operator. The density matrix and the interaction Hamiltonian in the interaction picture are defined by
\be
\rho^{(\text{int})}(t) &=& e^{i (H_S+H_E) t} \rho(t) e^{-i (H_S+H_E) t} \,, \\
H^{(\text{int})}_I(t) &=& e^{i (H_S+H_E) t} H_I e^{-i (H_S+H_E) t} = \int\diff x\, O^{(\text{int})}_E(t,x)
O^{(\text{int})}_S(t,x)\,,
\ee
respectively, where $O^{(\text{int})}_{E/S}(t,x) = e^{iH_{E/S}t} O_{E/S}(x) e^{-iH_{E/S}t}$. Since the environment scalar field is thermal, we have
\be
\rho_E^{(\text{int})}(t) = \rho_E = \frac{e^{-\beta H_E}}{\Tr_E e^{-\beta H_E}} \,,
\ee
where $\beta=1/T$ and $T$ is the temperature of the environment.
If we assume that the initial density matrix factorizes
\be
\label{eqn:factorize}
\rho^{(\text{int})}(t=0) = \rho_S^{(\text{int})}(t=0) \otimes \rho_E \,,
\ee
and that the interaction between the system and the environment is weak (the system and the environment themselves can be strongly-coupled), we obtain by expanding the evolution operator to second order in $H_I^{(\text{int})}$, the following result
\be
\label{eqn:pre-lindblad}
\rho_S^{(\text{int})}(t) &=& \rho_S^{(\text{int})}(0) 
- \int_0^t\diff t_1 \int_0^t\diff t_2 \int\diff x_1 \int\diff x_2 \frac{\sign(t_1-t_2)}{2}  D(t_1,x_1;t_2,x_2) \big[O^{(\text{int})}_S(t_1,x_1)O^{(\text{int})}_S(t_2,x_2), \rho_S^{(\text{int})}(0) \big]  \nn\\
&+& \int_0^t\diff t_1 \int_0^t\diff t_2 \int\diff x_1 \int\diff x_2 \, D(t_1,x_1;t_2,x_2) \Big( O^{(\text{int})}_S(t_2,x_2) \rho_S^{(\text{int})}(0) O^{(\text{int})}_S(t_1,x_1) \nn\\
&-& \frac{1}{2} \big\{O^{(\text{int})}_S(t_1,x_1)O^{(\text{int})}_S(t_2,x_2), \rho_S^{(\text{int})}(0) \big\} \Big) + \ml{O}\big((tH^{(\text{int})}_I)^3\big) \,.
\ee
Here the environment correlator is defined as
\be
D(t_1,x_1;t_2,x_2) = \Tr_E \big( \rho_E\, O^{(\text{int})}_E(t_1,x_1) O^{(\text{int})}_E(t_2,x_2) \big)\,,
\ee
and we have used $\Tr_E \big( \rho_E\, O^{(\text{int})}_E(t,x) \big)=0$, which can be realized by redefinitions of $O_E$ and $H_S$~\cite{Yao:2021lus}.

The expression in Eq.\ (\ref{eqn:pre-lindblad}) is a finite-difference equation. It can be converted into a well-defined differential equation in the quantum Brownian motion limit. The limit of quantum Brownian motion is specified by the following separation of time scales:
\be
\tau_R &\gg& \tau_E \,,\\
\tau_S &\gg& \tau_E \,,
\ee
where $\tau_R$ is the relaxation time of the system, $\tau_E$ denotes the environment correlation time and $\tau_S$ represents the intrinsic time scale of the system. The hierarchy $\tau_R \gg \tau_E$ leads to Markovian dynamics and is generally true when the interaction described by $H_I$ is weak. The hierarchy $\tau_S \gg \tau_E$ is valid if $T\gg H_S$ since $\tau_E\sim1/T$ and $\tau_S\sim1/H_S$. Simplifying Eq.\ (\ref{eqn:pre-lindblad}) using these two hierarchies of time scales leads to the Sch\"odinger-picture Lindblad equation in the limit of quantum Brownian motion (Some higher order terms in the expansion of $\frac{\tau_E}{\tau_S}$ are included here. Details of the derivation can be found in Ref.~\cite{Yao:2021lus}.)
\be
\label{eq:Lindblad}
\frac{\diff \rho_S(t)}{\diff t} = -i \big[H_S+\Delta H_S,  \rho_S(t) \big] + \int\diff x_1 \diff x_2 \, D(k_0=0, {x}_1-{x}_2)  \Big( \widetilde{O}_S({x}_2) \rho_S(t) \widetilde{O}^\dagger_S({x}_1) 
- \frac{1}{2} \big\{\widetilde{O}^\dagger_S({x}_1) \widetilde{O}_S({x}_2), \rho_S(t) \big\} \Big)\,. \nn\\
\ee
Here $\Delta H_S$ denotes the correction to the system Hamiltonian due to the interaction with the environment and we have
\begin{align}
\Delta H_S &\equiv \frac{1}{2}\int\diff x_1\diff x_2\, \Sigma(k_0=0,x_1-x_2) O_S(x_1) O_S(x_2)  \nn\\ &+ \frac{1}{4}\int\diff x_1\diff x_2\, \frac{\partial \Sigma(k_0=0,x_1-x_2)}{\partial k_0} \Big( \big[ H_S, O_S(x_1) \big] O_S(x_2) -  O_S(x_1) \big[ H_S, O_S(x_2) \big] \Big) \,,\\
\label{eqn:SigmaD}
\Sigma(k_0,x_1-x_2) &\equiv -i \int\diff (t_1-t_2) e^{ik_0(t_1-t_2)} \sign(t_1-t_2) D(t_1,x_1;t_2,x_2) \,, \\
\label{eqn:D}
D(k_0,x_1-x_2) &\equiv \int\diff (t_1-t_2) e^{ik_0(t_1-t_2)} D(t_1,x_1;t_2,x_2) \,, \\
\widetilde{O}_S(x) &\equiv O_S(x) - \frac{1}{4T} \big[H_S, O_S(x) \big] \,, \\
\widetilde{O}_S^\dagger(x) &\equiv O_S(x) + \frac{1}{4T} \big[H_S, O_S(x) \big] \,.
\end{align}
When defining the Fourier transform of the environment correlators in Eqs.~(\ref{eqn:SigmaD}) and (\ref{eqn:D}), we have assumed that the environment is invariant under spacetime translations. We drop the correction to the system Hamiltonian $\Delta H_S$ in the main text since we want the equilibrium state of the evolution equation to be the thermal state of the Schwinger model in the vacuum. It is necessary to keep the commutator terms in the definitions of $\widetilde{O}_S$ and $\widetilde{O}^\dagger_S$ for the system to thermalize (approximately).

For the Hilbert space of the zero-momentum states in the Schwinger model, considered in the main text, only the environment correlator $D(k_0=0,k=0)$ contributes. To see this more explicitly, we sandwich Eq.\ (\ref{eq:Lindblad}) with $\langle k=0,\alpha|$ and $|k=0,\beta\rangle$ and insert the identity $\sum_{\alpha}|k=0, \alpha \rangle \langle k=0,\alpha|=I$ (this is complete since we constrain the Hilbert space to include just the zero-momentum states) where the quantum numbers $\alpha$ and $\beta$ label different zero-momentum states. Denoting $\langle k=0,\alpha| \rho_S(t) |k=0,\beta\rangle = \rho_S^{\alpha\beta}(t)$, $\langle k=0,\alpha|H_S|k=0,\beta\rangle = H_S^{\alpha\beta} $, and $\langle k=0,\alpha|\widetilde{O}_S(x)|k=0,\beta\rangle = \widetilde{O}_S^{\alpha\beta}(x)$, we obtain
\be
\frac{\diff \rho_S^{\alpha\beta}(t) }{\diff t} &=& -i\sum_{\gamma} \big( H_S^{\alpha\gamma} \rho_S^{\gamma\beta}(t) - \rho_S^{\alpha\gamma}(t) H_S^{\gamma\beta} \big) + \int\diff x_1 \diff x_2\, D(k_0=0,x_1-x_2) \nn\\
&&\times
\Big(\widetilde{O}_S^{\alpha\gamma}(x_2) \rho_S^{\gamma\delta}(t)
\widetilde{O}_S^{\dagger\delta\beta}(x_1) - \frac{1}{2} \widetilde{O}_S^{\dagger\alpha\gamma}(x_1) \widetilde{O}_S^{\gamma\delta}(x_2) \rho_S^{\delta\beta}(t) - \frac{1}{2} \rho_S^{\alpha\gamma}(t)
\widetilde{O}_S^{\dagger\gamma\delta}(x_1) \widetilde{O}_S^{\delta\beta}(x_2)
\Big) \,.
\ee
Since the basis states have zero momentum, the projection of the system operator $O_S(x)$,
\be
\langle k=0,\alpha|\widetilde{O}_S(x)|k=0,\beta\rangle &=& \langle k=0,\alpha| e^{-i\hat{p}x} \widetilde{O}_S(x=0) e^{i\hat{p}x} |k=0,\beta\rangle = \langle k=0,\alpha|  \widetilde{O}_S(x=0) |k=0,\beta\rangle \nn\\
&=& \frac{1}{2Na}\int\diff x\, \langle k=0,\alpha|  \widetilde{O}_S(x) |k=0,\beta\rangle
\,,
\ee
is independent of the position $x$, where $\hat{p}$ is the momentum operator. Therefore, we can drop the dependence on $x_1$ and $x_2$ of the operators $\widetilde{O}_S$ and $\widetilde{O}^\dagger_S$, and we obtain
\be
\label{eqn:lindblad_matrix}
\frac{\diff \rho_S^{\alpha\beta}(t) }{\diff t} &=& -i\sum_{\gamma} \big( H_S^{\alpha\gamma} \rho_S^{\gamma\beta}(t) - \rho_S^{\alpha\gamma}(t) H_S^{\gamma\beta} \big) + \int\diff x\, D(k_0=0,k=0) \nn\\
&&\times
\Big(\widetilde{O}_S^{\alpha\gamma} \rho_S^{\gamma\delta}(t)
\widetilde{O}_S^{\dagger\delta\beta} - \frac{1}{2} \widetilde{O}_S^{\dagger\alpha\gamma} \widetilde{O}_S^{\gamma\delta} \rho_S^{\delta\beta}(t) - \frac{1}{2} \rho_S^{\alpha\gamma}(t)
\widetilde{O}_S^{\dagger\gamma\delta} \widetilde{O}_S^{\delta\beta}
\Big) \,.
\ee 
Here $\int\diff x\,D(k_0=0,k=0)$ is independent of the position. Different values of the environment correlator $D(k_0=0,k=0)$ will only modify the rate at which the system approaches equilibrium. The equilibrium properties of the system are independent of $D(k_0=0,k=0)$. Therefore, we will take the constant $D(k_0=0,k=0)$ as an input parameter in our calculations. The mass dimension of $D(k_0=0,k=0)$ is $0$. The matrix representation of the system operator $O_S^{\alpha\beta}$ in the discretized Schwinger model can be written as
\be
O_S^{\alpha\beta} &=& \frac{1}{aN_f}\int\diff x\, \langle k=0,\alpha|  \widetilde{O}_S(x) |k=0,\beta\rangle =  \frac{1}{aN_f}\int\diff x\, \langle k=0,\alpha|  \overline{\psi}\psi(x) |k=0,\beta\rangle \nn\\
&=& \frac{1}{aN_f}\sum_n \,  \langle k=0,\alpha| (-1)^n \overline{\chi}\chi(n) |k=0,\beta\rangle \nn\\
&=& \frac{1}{aN_f}\sum_n \,  \langle k=0,\alpha| (-1)^n \big( \overline{\chi}_u\chi_u(n) + \overline{\chi}_d\chi_d(n) \big) |k=0,\beta\rangle \nn\\
&=& \frac{1}{aN_f} \sum_n \,  \Big\langle k=0,\alpha \Big| \frac{(-1)^n (\sigma_z(n)+1)}{2}  \Big| k=0,\beta \Big\rangle  \,,
\ee
where we have dropped the lower component $\chi_d$ as before.
Defining the Lindblad operator as
\be
L = \sqrt{a{N_f} D(k_0=0,k=0)} \Big( O_S - \frac{1}{4T}\big[ H_S, O_S \big]
\Big)\,,
\ee
we can rewrite the Lindblad equation in Eq.\ (\ref{eqn:lindblad_matrix}) in the form
\be
\frac{\diff \rho_S(t)}{\diff t} = -i \big[H_S,\rho_S(t) \big] + L\rho_S(t) L^\dagger - \frac{1}{2}\big\{ L^\dagger L ,\rho_S(t) \big\}\,,
\ee
where we have omitted the matrix indices.

\section{Implementation on IBMQ}
\label{app:e}
\begin{figure}[t!]
\includegraphics[scale=0.6]{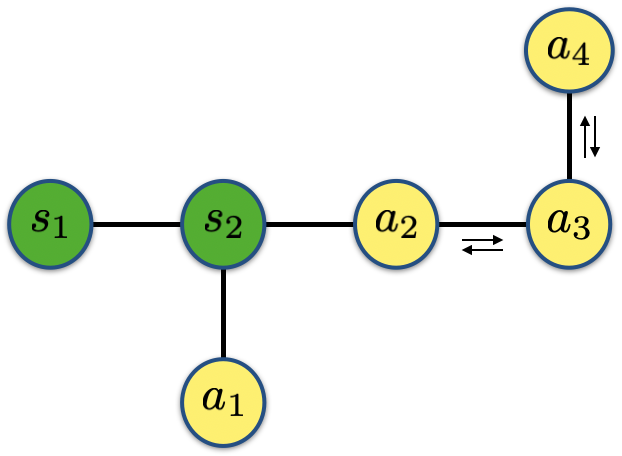}
\caption{Configuration of system $s_{1,2}$ and ancilla $a_{1-4}$ qubits on IBMQ. For the 3rd (4th) cycle/ancilla qubit we use 1 (2) additional swap operations.~\label{fig:system_ancilla_qubits}}
\end{figure}

In order to simulate the Lindblad equation of the Schwinger model in Eq.~(\ref{eqn:lindblad_maintext}), we first identify the time range where a given number of cycles (see Fig.~\ref{fig:algorithm}) yields a good approximation of the full RK4 result using the IBM's simulator. For example, we find that up to $t=0.5$, 1 cycle is sufficient for the parameters chosen in the main text, Fig.~\ref{fig:device}. For larger values of $t$, we switch to 2 cycles and eventually up to 3 and 4 cycles. For each time interval we use an appropriate optimization threshold in \texttt{qsearch}~\cite{2020arXiv201000215H}. The optimization assumes a linear qubit topology with only nearest-neighbor CNOT gates. We verify again with the simulator that the chosen thresholds give an approximation of the RK4 solution within a few percent, i.e. within the error that we can currently achieve on the quantum chip. For 4 cycles we obtain up to $\approx 50$ CNOT gates. Despite this large number we obtain good results from the \texttt{ibmq}$\bf{\_}$\texttt{montreal} device~\cite{IBMQMontreal} even without applying additional error mitigation techniques. We also verified that we obtain similar results from the \texttt{ibmq}$\bf{\_}$\texttt{toronto} device~\cite{IBMQToronto}. 
We ran each circuit for $15\times 8192$ shots since the CNOT gate error mitigation using the Random Identity Insertion Method of Ref.~\cite{PhysRevA.102.012426} requires high statistics.

The relevant part of the qubit topology which we use on the \texttt{ibmq}$\bf{\_}$\texttt{montreal} device~\cite{IBMQMontreal} is illustrated in Fig.~\ref{fig:system_ancilla_qubits}, where the system and ancilla qubits are highlighted with different colors. As mentioned in the main text, we use a different ancilla qubit for every cycle. In our setup the ancilla qubits need to be connected to the system qubit $s_2$ in Fig.~\ref{fig:system_ancilla_qubits}. In order to run 3 (4) cycles, we make use of 1 (2) additional swap operations of the ancilla qubits,
each of which consist of 3 CNOT gates.
These CNOT gates are included in the CNOT gate error mitigation procedure.

\section{Validation of the quantum circuit for $N=4$}
\label{app:f}
\begin{figure}[!h]
\includegraphics[scale=0.53]{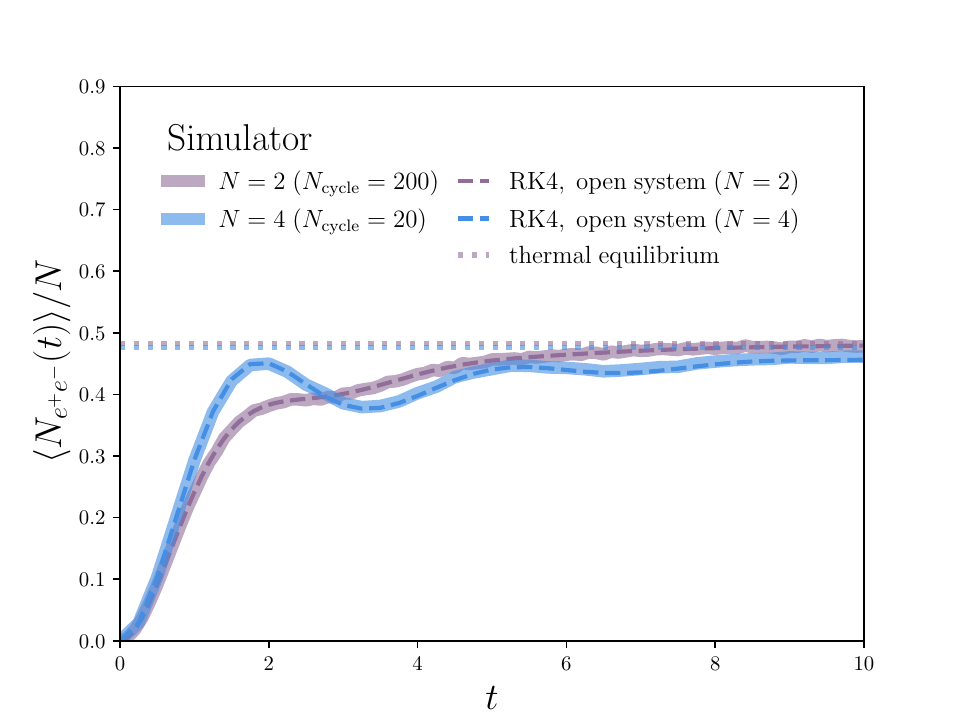}
\includegraphics[scale=0.53]{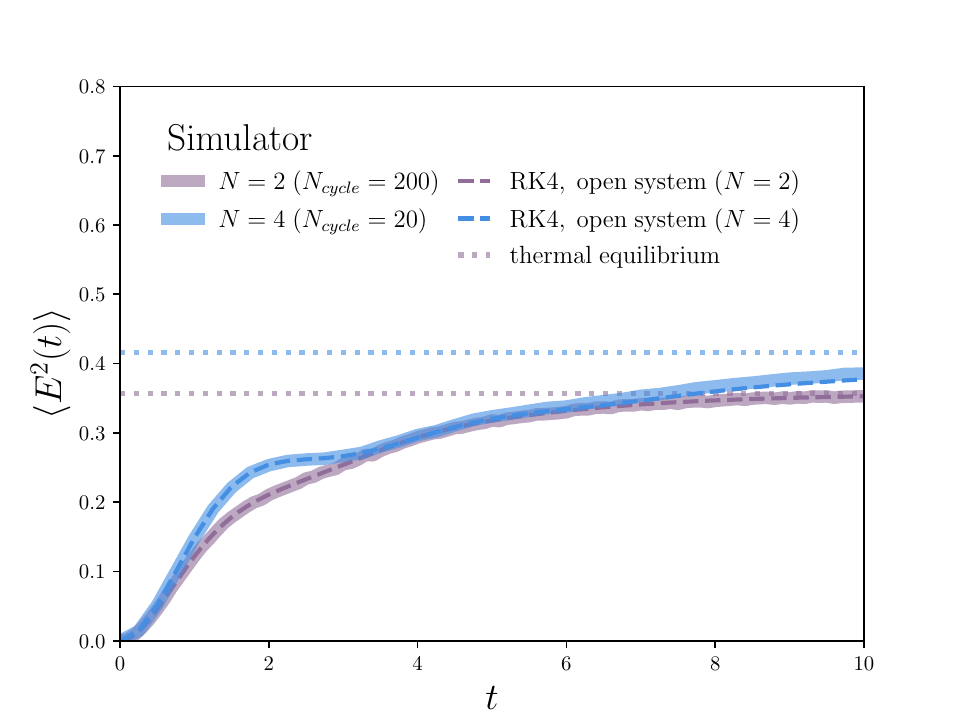}
\caption{Quantum simulation of non-equilibrium dynamics in the Schwinger model: $\langle N_{e^+e^-} \rangle$ and $\langle E^2 \rangle$ using the quantum circuit for $N=2$ and $N=4$, along with numerical solution (RK4). 
The same values of the parameters are used as in Fig.~\ref{fig:simulation-N-dependence-E2}.
~\label{fig:simulation-N4}}
\end{figure}

In order to further validate the performance of the quantum circuit
beyond $N=2$,
we simulate the quantum circuit using the IBMQ (noiseless) simulator
for $N=4$. Fig.~\ref{fig:simulation-N4} shows the result
along with comparison to the corresponding numerical solutions,
which show good agreement.

\section{Volume-dependence for $\left<N_{\mathrm{e^+e^-}}\right>$}
\label{app:g}
\begin{figure}[!h]
\includegraphics[scale=0.53]{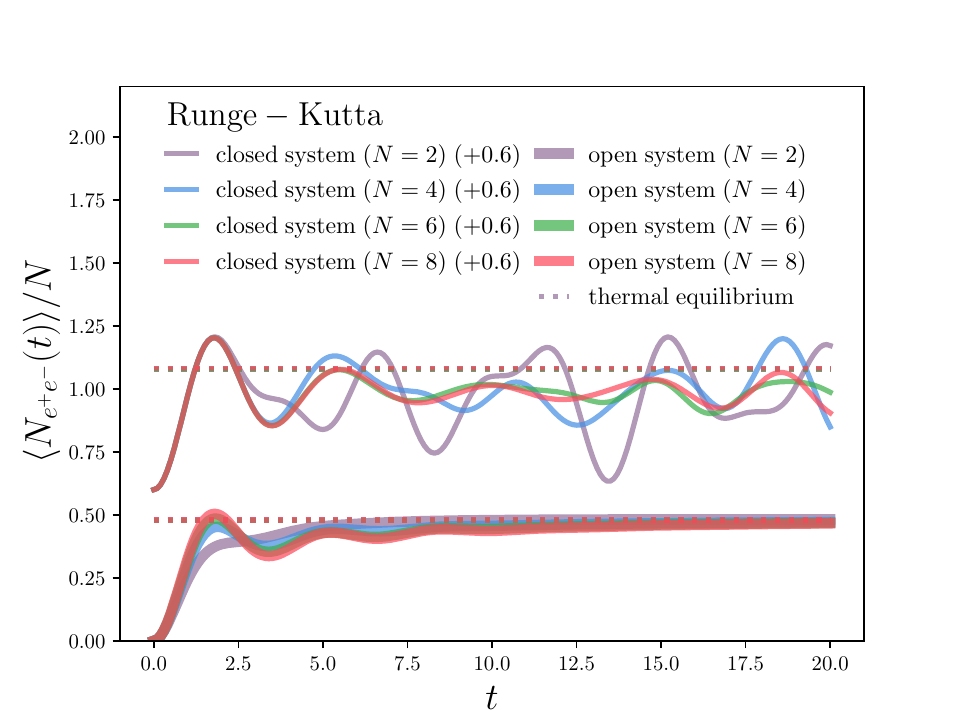}
\caption{Numerical solution of non-equilibrium dynamics in the Schwinger model: $\langle N_{e^+e^-} \rangle$ up to $N=8$. 
The same values of the parameters are used as in Fig.~\ref{fig:simulation-N-dependence-E2}.
~\label{fig:simulation-N-dependence-Ne+e-}}
\end{figure}

In order to accompany the volume-dependence of $\left< E^2 \right>$ 
shown in in Fig.~\ref{fig:simulation-N-dependence-E2}, we plot
the average number of electron-positron pairs as a function of 
$N$ for fixed lattice spacing $a=1$ in Fig.~\ref{fig:simulation-N-dependence-Ne+e-}. 

\bibliographystyle{utphys}
\bibliography{main.bib}

\providecommand{\href}[2]{#2}\begingroup\raggedright\begin{thebibliography}{100}

\bibitem{Devoret2013}
M.~Devoret and R.~Schoelkopf, ``Superconducting circuits for quantum
  information: An outlook,''
  \href{http://dx.doi.org/10.1126/science.1231930}{{\em Science (New York,
  N.Y.)} {\bfseries 339} (03, 2013) 1169--74}.

\bibitem{annurev-conmatphys-031119-050605}
M.~Kjaergaard, M.~E. Schwartz, J.~Braumüller, P.~Krantz, J.~I.-J. Wang,
  S.~Gustavsson, and W.~D. Oliver, ``Superconducting qubits: Current state of
  play,''
  \href{http://dx.doi.org/10.1146/annurev-conmatphys-031119-050605}{{\em Annual
  Review of Condensed Matter Physics} {\bfseries 11} no.~1, (2020) 369--395}.

\bibitem{doi:10.1063/1.5088164}
C.~D. Bruzewicz, J.~Chiaverini, R.~McConnell, and J.~M. Sage, ``Trapped-ion
  quantum computing: Progress and challenges,''
  \href{http://dx.doi.org/10.1063/1.5088164}{{\em Applied Physics Reviews}
  {\bfseries 6} no.~2, (2019) 021314}.

\bibitem{google_supremacy}
F.~Arute, K.~Arya, R.~Babbush, D.~Bacon, J.~C. Bardin, R.~Barends, R.~Biswas,
  S.~Boixo, F.~G. S.~L. Brandao, and D.~A. e.~a. Buell, ``Quantum supremacy
  using a programmable superconducting processor,''
  \href{http://dx.doi.org/10.1038/s41586-019-1666-5}{{\em Nature} {\bfseries
  574} no.~7779, (2019) 505--510}.

\bibitem{PhysRevB.101.184305}
L.~Bassman, K.~Liu, A.~Krishnamoorthy, T.~Linker, Y.~Geng, D.~Shebib,
  S.~Fukushima, F.~Shimojo, R.~K. Kalia, A.~Nakano, and P.~Vashishta, ``Towards
  simulation of the dynamics of materials on quantum computers,''
  \href{http://dx.doi.org/10.1103/PhysRevB.101.184305}{{\em Phys. Rev. B}
  {\bfseries 101} (May, 2020) 184305}.

\bibitem{Smith2019}
A.~Smith, M.~S. Kim, F.~Pollmann, and J.~Knolle, ``Simulating quantum many-body
  dynamics on a current digital quantum computer,''
  \href{http://dx.doi.org/10.1038/s41534-019-0217-0}{{\em npj Quantum
  Information} {\bfseries 5} (11, 2019) 106}.

\bibitem{Preskill_2018}
J.~Preskill, ``Quantum computing in the nisq era and beyond,''
  \href{http://dx.doi.org/10.22331/q-2018-08-06-79}{{\em Quantum} {\bfseries 2}
  (Aug, 2018) 79}.

\bibitem{Kaplan:2017ccd}
D.~B. Kaplan, N.~Klco, and A.~Roggero, ``{Ground States via Spectral Combing on
  a Quantum Computer},'' \href{http://arxiv.org/abs/1709.08250}{{\ttfamily
  arXiv:1709.08250 [quant-ph]}}.

\bibitem{Preskill:2018fag}
J.~Preskill, ``{Simulating quantum field theory with a quantum computer},''
  \href{http://dx.doi.org/10.22323/1.334.0024}{{\em PoS} {\bfseries
  LATTICE2018} (2018) 024}, \href{http://arxiv.org/abs/1811.10085}{{\ttfamily
  arXiv:1811.10085 [hep-lat]}}.

\bibitem{Lamm:2018siq}
H.~Lamm and S.~Lawrence, ``{Simulation of Nonequilibrium Dynamics on a Quantum
  Computer},'' \href{http://dx.doi.org/10.1103/PhysRevLett.121.170501}{{\em
  Phys. Rev. Lett.} {\bfseries 121} no.~17, (2018) 170501},
  \href{http://arxiv.org/abs/1806.06649}{{\ttfamily arXiv:1806.06649
  [quant-ph]}}.

\bibitem{Dumitrescu:2018njn}
E.~Dumitrescu, A.~McCaskey, G.~Hagen, G.~Jansen, T.~Morris, T.~Papenbrock,
  R.~Pooser, D.~Dean, and P.~Lougovski, ``{Cloud Quantum Computing of an Atomic
  Nucleus},'' \href{http://dx.doi.org/10.1103/PhysRevLett.120.210501}{{\em
  Phys. Rev. Lett.} {\bfseries 120} no.~21, (2018) 210501},
  \href{http://arxiv.org/abs/1801.03897}{{\ttfamily arXiv:1801.03897
  [quant-ph]}}.

\bibitem{Roggero:2019myu}
A.~Roggero, A.~C. Li, J.~Carlson, R.~Gupta, and G.~N. Perdue, ``{Quantum
  Computing for Neutrino-Nucleus Scattering},''
  \href{http://dx.doi.org/10.1103/PhysRevD.101.074038}{{\em Phys. Rev. D}
  {\bfseries 101} no.~7, (2020) 074038},
  \href{http://arxiv.org/abs/1911.06368}{{\ttfamily arXiv:1911.06368
  [quant-ph]}}.

\bibitem{Bauer:2019qxa}
C.~W. Bauer, W.~A. de~Jong, B.~Nachman, and D.~Provasoli, ``{Quantum Algorithm
  for High Energy Physics Simulations},''
  \href{http://dx.doi.org/10.1103/PhysRevLett.126.062001}{{\em Phys. Rev.
  Lett.} {\bfseries 126} no.~6, (2021) 062001},
  \href{http://arxiv.org/abs/1904.03196}{{\ttfamily arXiv:1904.03196
  [hep-ph]}}.

\bibitem{Mueller:2019qqj}
N.~Mueller, A.~Tarasov, and R.~Venugopalan, ``{Deeply inelastic scattering
  structure functions on a hybrid quantum computer},''
  \href{http://dx.doi.org/10.1103/PhysRevD.102.016007}{{\em Phys. Rev. D}
  {\bfseries 102} no.~1, (2020) 016007},
  \href{http://arxiv.org/abs/1908.07051}{{\ttfamily arXiv:1908.07051
  [hep-th]}}.

\bibitem{Wei:2019rqy}
A.~Y. Wei, P.~Naik, A.~W. Harrow, and J.~Thaler, ``{Quantum Algorithms for Jet
  Clustering},'' \href{http://dx.doi.org/10.1103/PhysRevD.101.094015}{{\em
  Phys. Rev. D} {\bfseries 101} no.~9, (2020) 094015},
  \href{http://arxiv.org/abs/1908.08949}{{\ttfamily arXiv:1908.08949
  [hep-ph]}}.

\bibitem{Holland:2019zju}
E.~T. Holland, K.~A. Wendt, K.~Kravvaris, X.~Wu, W.~Erich~Ormand, J.~L. DuBois,
  S.~Quaglioni, and F.~Pederiva, ``{Optimal Control for the Quantum Simulation
  of Nuclear Dynamics},''
  \href{http://dx.doi.org/10.1103/PhysRevA.101.062307}{{\em Phys. Rev. A}
  {\bfseries 101} no.~6, (2020) 062307},
  \href{http://arxiv.org/abs/1908.08222}{{\ttfamily arXiv:1908.08222
  [quant-ph]}}.

\bibitem{Klco:2019evd}
N.~Klco, J.~R. Stryker, and M.~J. Savage, ``{SU(2) non-Abelian gauge field
  theory in one dimension on digital quantum computers},''
  \href{http://dx.doi.org/10.1103/PhysRevD.101.074512}{{\em Phys. Rev. D}
  {\bfseries 101} no.~7, (2020) 074512},
  \href{http://arxiv.org/abs/1908.06935}{{\ttfamily arXiv:1908.06935
  [quant-ph]}}.

\bibitem{Avkhadiev:2019niu}
A.~Avkhadiev, P.~Shanahan, and R.~Young, ``{Accelerating Lattice Quantum Field
  Theory Calculations via Interpolator Optimization Using Noisy
  Intermediate-Scale Quantum Computing},''
  \href{http://dx.doi.org/10.1103/PhysRevLett.124.080501}{{\em Phys. Rev.
  Lett.} {\bfseries 124} no.~8, (2020) 080501},
  \href{http://arxiv.org/abs/1908.04194}{{\ttfamily arXiv:1908.04194
  [hep-lat]}}.

\bibitem{DeJong:2020riy}
W.~A. De~Jong, M.~Metcalf, J.~Mulligan, M.~P\l{}osko\'n, F.~Ringer, and X.~Yao,
  ``{Quantum simulation of open quantum systems in heavy-ion collisions},''
  \href{http://dx.doi.org/10.1103/PhysRevD.104.L051501}{{\em Phys. Rev. D}
  {\bfseries 104} no.~5, (2021) 051501},
  \href{http://arxiv.org/abs/2010.03571}{{\ttfamily arXiv:2010.03571
  [hep-ph]}}.

\bibitem{Liu:2020eoa}
J.~Liu and Y.~Xin, ``{Quantum simulation of quantum field theories as quantum
  chemistry},'' \href{http://dx.doi.org/10.1007/JHEP12(2020)011}{{\em JHEP}
  {\bfseries 12} (2020) 011}, \href{http://arxiv.org/abs/2004.13234}{{\ttfamily
  arXiv:2004.13234 [hep-th]}}.

\bibitem{Kreshchuk:2020dla}
M.~Kreshchuk, W.~M. Kirby, G.~Goldstein, H.~Beauchemin, and P.~J. Love,
  ``{Quantum simulation of quantum field theory in the light-front
  formulation},'' \href{http://dx.doi.org/10.1103/PhysRevA.105.032418}{{\em
  Phys. Rev. A} {\bfseries 105} no.~3, (2022) 032418},
  \href{http://arxiv.org/abs/2002.04016}{{\ttfamily arXiv:2002.04016
  [quant-ph]}}.

\bibitem{Davoudi:2020yln}
Z.~Davoudi, I.~Raychowdhury, and A.~Shaw, ``{Search for efficient formulations
  for Hamiltonian simulation of non-Abelian lattice gauge theories},''
  \href{http://dx.doi.org/10.1103/PhysRevD.104.074505}{{\em Phys. Rev. D}
  {\bfseries 104} no.~7, (2021) 074505},
  \href{http://arxiv.org/abs/2009.11802}{{\ttfamily arXiv:2009.11802
  [hep-lat]}}.

\bibitem{Briceno:2020rar}
R.~A. Brice\~no, J.~V. Guerrero, M.~T. Hansen, and A.~M. Sturzu, ``{Role of
  boundary conditions in quantum computations of scattering observables},''
  \href{http://dx.doi.org/10.1103/PhysRevD.103.014506}{{\em Phys. Rev. D}
  {\bfseries 103} no.~1, (2021) 014506},
  \href{http://arxiv.org/abs/2007.01155}{{\ttfamily arXiv:2007.01155
  [hep-lat]}}.

\bibitem{Echevarria:2020wct}
M.~G. Echevarria, I.~L. Egusquiza, E.~Rico, and G.~Schnell, ``{Quantum
  simulation of light-front parton correlators},''
  \href{http://dx.doi.org/10.1103/PhysRevD.104.014512}{{\em Phys. Rev. D}
  {\bfseries 104} no.~1, (2021) 014512},
  \href{http://arxiv.org/abs/2011.01275}{{\ttfamily arXiv:2011.01275
  [quant-ph]}}.

\bibitem{Chang:2020iwh}
C.~C. Chang, C.-C. Chen, C.~Koerber, T.~S. Humble, and J.~Ostrowski, ``{Integer
  Programming from Quantum Annealing and Open Quantum Systems},''
  \href{http://arxiv.org/abs/2009.11970}{{\ttfamily arXiv:2009.11970
  [quant-ph]}}.

\bibitem{Hubisz:2020vhx}
J.~Hubisz, B.~Sambasivam, and J.~Unmuth-Yockey, ``{Quantum algorithms for open
  lattice field theory},''
  \href{http://dx.doi.org/10.1103/PhysRevA.104.052420}{{\em Phys. Rev. A}
  {\bfseries 104} no.~5, (2021) 052420},
  \href{http://arxiv.org/abs/2012.05257}{{\ttfamily arXiv:2012.05257
  [hep-lat]}}.

\bibitem{Cohen:2021imf}
{\bfseries NuQS} Collaboration, T.~D. Cohen, H.~Lamm, S.~Lawrence, and
  Y.~Yamauchi, ``{Quantum algorithms for transport coefficients in gauge
  theories},'' \href{http://dx.doi.org/10.1103/PhysRevD.104.094514}{{\em Phys.
  Rev. D} {\bfseries 104} no.~9, (2021) 094514},
  \href{http://arxiv.org/abs/2104.02024}{{\ttfamily arXiv:2104.02024
  [hep-lat]}}.

\bibitem{Barata:2021yri}
J.~a. Barata and C.~A. Salgado, ``{A quantum strategy to compute the jet
  quenching parameter $\hat{q}$},''
  \href{http://arxiv.org/abs/2104.04661}{{\ttfamily arXiv:2104.04661
  [hep-ph]}}.

\bibitem{Ramirez-Uribe:2021ubp}
S.~Ram\'\i{}rez-Uribe, A.~E. Renter\'\i{}a-Olivo, G.~Rodrigo, G.~F.~R.
  Sborlini, and L.~Vale~Silva, ``{Quantum algorithm for Feynman loop
  integrals},'' \href{http://dx.doi.org/10.1007/JHEP05(2022)100}{{\em JHEP}
  {\bfseries 05} (2022) 100}, \href{http://arxiv.org/abs/2105.08703}{{\ttfamily
  arXiv:2105.08703 [hep-ph]}}.

\bibitem{Li:2021kcs}
{\bfseries QuNu} Collaboration, T.~Li, X.~Guo, W.~K. Lai, X.~Liu, E.~Wang,
  H.~Xing, D.-B. Zhang, and S.-L. Zhu, ``{Partonic collinear structure by
  quantum computing},''
  \href{http://dx.doi.org/10.1103/PhysRevD.105.L111502}{{\em Phys. Rev. D}
  {\bfseries 105} no.~11, (2022) L111502},
  \href{http://arxiv.org/abs/2106.03865}{{\ttfamily arXiv:2106.03865
  [hep-ph]}}.

\bibitem{Kogut:1974ag}
J.~B. Kogut and L.~Susskind, ``{Hamiltonian Formulation of Wilson's Lattice
  Gauge Theories},'' \href{http://dx.doi.org/10.1103/PhysRevD.11.395}{{\em
  Phys. Rev. D} {\bfseries 11} (1975) 395--408}.

\bibitem{Jordan:2011ci}
S.~P. Jordan, K.~S.~M. Lee, and J.~Preskill, ``{Quantum Computation of
  Scattering in Scalar Quantum Field Theories},'' {\em Quant. Inf. Comput.}
  {\bfseries 14} (2014) 1014--1080,
  \href{http://arxiv.org/abs/1112.4833}{{\ttfamily arXiv:1112.4833 [hep-th]}}.

\bibitem{Jordan:2011ne}
S.~P. Jordan, K.~S. Lee, and J.~Preskill, ``{Quantum Algorithms for Quantum
  Field Theories},'' \href{http://dx.doi.org/10.1126/science.1217069}{{\em
  Science} {\bfseries 336} (2012) 1130--1133},
  \href{http://arxiv.org/abs/1111.3633}{{\ttfamily arXiv:1111.3633
  [quant-ph]}}.

\bibitem{Jordan:2014tma}
S.~P. Jordan, K.~S.~M. Lee, and J.~Preskill, ``{Quantum Algorithms for
  Fermionic Quantum Field Theories},''
  \href{http://arxiv.org/abs/1404.7115}{{\ttfamily arXiv:1404.7115 [hep-th]}}.

\bibitem{Jordan:2017lea}
S.~P. Jordan, H.~Krovi, K.~S.~M. Lee, and J.~Preskill, ``{BQP-completeness of
  Scattering in Scalar Quantum Field Theory},''
  \href{http://dx.doi.org/10.22331/q-2018-01-08-44}{{\em Quantum} {\bfseries 2}
  (2018) 44}, \href{http://arxiv.org/abs/1703.00454}{{\ttfamily
  arXiv:1703.00454 [quant-ph]}}.

\bibitem{Horn:1981kk}
D.~Horn, ``{Finite Matrix Models With Continuous Local Gauge Invariance},''
  \href{http://dx.doi.org/10.1016/0370-2693(81)90763-2}{{\em Phys. Lett. B}
  {\bfseries 100} (1981) 149--151}.

\bibitem{Orland:1989st}
P.~Orland and D.~Rohrlich, ``{Lattice Gauge Magnets: Local Isospin From
  Spin},'' \href{http://dx.doi.org/10.1016/0550-3213(90)90646-U}{{\em Nucl.
  Phys. B} {\bfseries 338} (1990) 647--672}.

\bibitem{Chandrasekharan:1996ih}
S.~Chandrasekharan and U.~J. Wiese, ``{Quantum link models: A Discrete approach
  to gauge theories},''
  \href{http://dx.doi.org/10.1016/S0550-3213(97)00006-0}{{\em Nucl. Phys. B}
  {\bfseries 492} (1997) 455--474},
  \href{http://arxiv.org/abs/hep-lat/9609042}{{\ttfamily
  arXiv:hep-lat/9609042}}.

\bibitem{Wiese:2013uua}
U.-J. Wiese, ``{Ultracold Quantum Gases and Lattice Systems: Quantum Simulation
  of Lattice Gauge Theories},''
  \href{http://dx.doi.org/10.1002/andp.201300104}{{\em Annalen Phys.}
  {\bfseries 525} (2013) 777--796},
  \href{http://arxiv.org/abs/1305.1602}{{\ttfamily arXiv:1305.1602
  [quant-ph]}}.

\bibitem{Muschik:2016tws}
C.~Muschik, M.~Heyl, E.~Martinez, T.~Monz, P.~Schindler, B.~Vogell,
  M.~Dalmonte, P.~Hauke, R.~Blatt, and P.~Zoller, ``{U(1) Wilson lattice gauge
  theories in digital quantum simulators},''
  \href{http://dx.doi.org/10.1088/1367-2630/aa89ab}{{\em New J. Phys.}
  {\bfseries 19} no.~10, (2017) 103020},
  \href{http://arxiv.org/abs/1612.08653}{{\ttfamily arXiv:1612.08653
  [quant-ph]}}.

\bibitem{Martinez:2016yna}
E.~A. Martinez {\em et~al.}, ``{Real-time dynamics of lattice gauge theories
  with a few-qubit quantum computer},''
  \href{http://dx.doi.org/10.1038/nature18318}{{\em Nature} {\bfseries 534}
  (2016) 516--519}, \href{http://arxiv.org/abs/1605.04570}{{\ttfamily
  arXiv:1605.04570 [quant-ph]}}.

\bibitem{Ercolessi:2017jbi}
E.~Ercolessi, P.~Facchi, G.~Magnifico, S.~Pascazio, and F.~V. Pepe, ``{Phase
  Transitions in $Z_{n}$ Gauge Models: Towards Quantum Simulations of the
  Schwinger-Weyl QED},''
  \href{http://dx.doi.org/10.1103/PhysRevD.98.074503}{{\em Phys. Rev. D}
  {\bfseries 98} no.~7, (2018) 074503},
  \href{http://arxiv.org/abs/1705.11047}{{\ttfamily arXiv:1705.11047
  [quant-ph]}}.

\bibitem{Klco:2018kyo}
N.~Klco, E.~Dumitrescu, A.~McCaskey, T.~Morris, R.~Pooser, M.~Sanz, E.~Solano,
  P.~Lougovski, and M.~Savage, ``{Quantum-classical computation of Schwinger
  model dynamics using quantum computers},''
  \href{http://dx.doi.org/10.1103/PhysRevA.98.032331}{{\em Phys. Rev. A}
  {\bfseries 98} no.~3, (2018) 032331},
  \href{http://arxiv.org/abs/1803.03326}{{\ttfamily arXiv:1803.03326
  [quant-ph]}}.

\bibitem{Raychowdhury:2018osk}
I.~Raychowdhury and J.~R. Stryker, ``{Solving Gauss's Law on Digital Quantum
  Computers with Loop-String-Hadron Digitization},''
  \href{http://dx.doi.org/10.1103/PhysRevResearch.2.033039}{{\em Phys. Rev.
  Res.} {\bfseries 2} no.~3, (2020) 033039},
  \href{http://arxiv.org/abs/1812.07554}{{\ttfamily arXiv:1812.07554
  [hep-lat]}}.

\bibitem{Klco:2018zqz}
N.~Klco and M.~J. Savage, ``{Digitization of scalar fields for quantum
  computing},'' \href{http://dx.doi.org/10.1103/PhysRevA.99.052335}{{\em Phys.
  Rev. A} {\bfseries 99} no.~5, (2019) 052335},
  \href{http://arxiv.org/abs/1808.10378}{{\ttfamily arXiv:1808.10378
  [quant-ph]}}.

\bibitem{Magnifico:2019kyj}
G.~Magnifico, M.~Dalmonte, P.~Facchi, S.~Pascazio, F.~V. Pepe, and
  E.~Ercolessi, ``{Real Time Dynamics and Confinement in the $\mathbb{Z}_{n}$
  Schwinger-Weyl lattice model for 1+1 QED},''
  \href{http://dx.doi.org/10.22331/q-2020-06-15-281}{{\em Quantum} {\bfseries
  4} (2020) 281}, \href{http://arxiv.org/abs/1909.04821}{{\ttfamily
  arXiv:1909.04821 [quant-ph]}}.

\bibitem{Chakraborty:2020uhf}
B.~Chakraborty, M.~Honda, T.~Izubuchi, Y.~Kikuchi, and A.~Tomiya, ``{Digital
  Quantum Simulation of the Schwinger Model with Topological Term via Adiabatic
  State Preparation},'' \href{http://arxiv.org/abs/2001.00485}{{\ttfamily
  arXiv:2001.00485 [hep-lat]}}.

\bibitem{Shaw:2020udc}
A.~F. Shaw, P.~Lougovski, J.~R. Stryker, and N.~Wiebe, ``{Quantum Algorithms
  for Simulating the Lattice Schwinger Model},''
  \href{http://dx.doi.org/10.22331/q-2020-08-10-306}{{\em Quantum} {\bfseries
  4} (2020) 306}, \href{http://arxiv.org/abs/2002.11146}{{\ttfamily
  arXiv:2002.11146 [quant-ph]}}.

\bibitem{Kharzeev:2020kgc}
D.~E. Kharzeev and Y.~Kikuchi, ``{Real-time chiral dynamics from a digital
  quantum simulation},''
  \href{http://dx.doi.org/10.1103/PhysRevResearch.2.023342}{{\em Phys. Rev.
  Res.} {\bfseries 2} no.~2, (2020) 023342},
  \href{http://arxiv.org/abs/2001.00698}{{\ttfamily arXiv:2001.00698
  [hep-ph]}}.

\bibitem{Ikeda:2020agk}
K.~Ikeda, D.~E. Kharzeev, and Y.~Kikuchi, ``{Real-time dynamics of Chern-Simons
  fluctuations near a critical point},''
  \href{http://dx.doi.org/10.1103/PhysRevD.103.L071502}{{\em Phys. Rev. D}
  {\bfseries 103} no.~7, (2021) L071502},
  \href{http://arxiv.org/abs/2012.02926}{{\ttfamily arXiv:2012.02926
  [hep-ph]}}.

\bibitem{Alexandru:2019nsa}
{\bfseries NuQS} Collaboration, A.~Alexandru, P.~F. Bedaque, S.~Harmalkar,
  H.~Lamm, S.~Lawrence, and N.~C. Warrington, ``{Gluon Field Digitization for
  Quantum Computers},''
  \href{http://dx.doi.org/10.1103/PhysRevD.100.114501}{{\em Phys. Rev. D}
  {\bfseries 100} no.~11, (2019) 114501},
  \href{http://arxiv.org/abs/1906.11213}{{\ttfamily arXiv:1906.11213
  [hep-lat]}}.

\bibitem{Barata:2020jtq}
J.~a. Barata, N.~Mueller, A.~Tarasov, and R.~Venugopalan, ``{Single-particle
  digitization strategy for quantum computation of a $\phi^4$ scalar field
  theory},'' \href{http://dx.doi.org/10.1103/PhysRevA.103.042410}{{\em Phys.
  Rev. A} {\bfseries 103} no.~4, (2021) 042410},
  \href{http://arxiv.org/abs/2012.00020}{{\ttfamily arXiv:2012.00020
  [hep-th]}}.

\bibitem{Harmalkar:2020mpd}
{\bfseries NuQS} Collaboration, S.~Harmalkar, H.~Lamm, and S.~Lawrence,
  ``{Quantum Simulation of Field Theories Without State Preparation},''
  \href{http://arxiv.org/abs/2001.11490}{{\ttfamily arXiv:2001.11490
  [hep-lat]}}.

\bibitem{Bauer:2021gup}
C.~W. Bauer, M.~Freytsis, and B.~Nachman, ``{Simulating Collider Physics on
  Quantum Computers Using Effective Field Theories},''
  \href{http://dx.doi.org/10.1103/PhysRevLett.127.212001}{{\em Phys. Rev.
  Lett.} {\bfseries 127} no.~21, (2021) 212001},
  \href{http://arxiv.org/abs/2102.05044}{{\ttfamily arXiv:2102.05044
  [hep-ph]}}.

\bibitem{Ciavarella:2021nmj}
A.~Ciavarella, N.~Klco, and M.~J. Savage, ``{Trailhead for quantum simulation
  of SU(3) Yang-Mills lattice gauge theory in the local multiplet basis},''
  \href{http://dx.doi.org/10.1103/PhysRevD.103.094501}{{\em Phys. Rev. D}
  {\bfseries 103} no.~9, (2021) 094501},
  \href{http://arxiv.org/abs/2101.10227}{{\ttfamily arXiv:2101.10227
  [quant-ph]}}.

\bibitem{Banuls:2019bmf}
M.~C. Ba\~nuls {\em et~al.}, ``{Simulating Lattice Gauge Theories within
  Quantum Technologies},''
  \href{http://dx.doi.org/10.1140/epjd/e2020-100571-8}{{\em Eur. Phys. J. D}
  {\bfseries 74} no.~8, (2020) 165},
  \href{http://arxiv.org/abs/1911.00003}{{\ttfamily arXiv:1911.00003
  [quant-ph]}}.

\bibitem{Cloet:2019wre}
I.~C. Cloët {\em et~al.}, ``{Opportunities for Nuclear Physics \& Quantum
  Information Science},'' in {\em {Intersections between Nuclear Physics and
  Quantum Information}}, I.~C. Cloët and M.~R. Dietrich, eds.
\newblock 3, 2019.
\newblock \href{http://arxiv.org/abs/1903.05453}{{\ttfamily arXiv:1903.05453
  [nucl-th]}}.

\bibitem{Zhang:2020uqo}
D.-B. Zhang, H.~Xing, H.~Yan, E.~Wang, and S.-L. Zhu, ``{Selected topics of
  quantum computing for nuclear physics},''
  \href{http://dx.doi.org/10.1088/1674-1056/abd761}{{\em Chin. Phys. B}
  {\bfseries 30} no.~2, (2021) 020306},
  \href{http://arxiv.org/abs/2011.01431}{{\ttfamily arXiv:2011.01431
  [quant-ph]}}.

\bibitem{temme2011quantum}
K.~Temme, T.~J. Osborne, K.~G. Vollbrecht, D.~Poulin, and F.~Verstraete,
  ``Quantum metropolis sampling,'' {\em Nature} {\bfseries 471} no.~7336,
  (2011) 87--90.

\bibitem{motta2020determining}
M.~Motta, C.~Sun, A.~T. Tan, M.~J. O’Rourke, E.~Ye, A.~J. Minnich, F.~G.
  Brandao, and G.~K.-L. Chan, ``Determining eigenstates and thermal states on a
  quantum computer using quantum imaginary time evolution,'' {\em Nature
  Physics} {\bfseries 16} no.~2, (2020) 205--210.

\bibitem{zalka1998simulating}
C.~Zalka, ``Simulating quantum systems on a quantum computer,'' {\em
  Proceedings of the Royal Society of London. Series A: Mathematical, Physical
  and Engineering Sciences} {\bfseries 454} no.~1969, (1998) 313--322.

\bibitem{terhal2000problem}
B.~M. Terhal and D.~P. DiVincenzo, ``Problem of equilibration and the
  computation of correlation functions on a quantum computer,'' {\em Physical
  Review A} {\bfseries 61} no.~2, (2000) 022301.

\bibitem{wang2011quantum}
H.~Wang, S.~Ashhab, and F.~Nori, ``Quantum algorithm for simulating the
  dynamics of an open quantum system,'' {\em Physical Review A} {\bfseries 83}
  no.~6, (2011) 062317.

\bibitem{PhysRevResearch.2.023214}
M.~Metcalf, J.~E. Moussa, W.~A. de~Jong, and M.~Sarovar, ``Engineered
  thermalization and cooling of quantum many-body systems,''
  \href{http://dx.doi.org/10.1103/PhysRevResearch.2.023214}{{\em Phys. Rev.
  Research} {\bfseries 2} (May, 2020) 023214}.

\bibitem{Young:2010jq}
C.~Young and K.~Dusling, ``{Quarkonium above deconfinement as an open quantum
  system},'' \href{http://dx.doi.org/10.1103/PhysRevC.87.065206}{{\em Phys.
  Rev. C} {\bfseries 87} no.~6, (2013) 065206},
  \href{http://arxiv.org/abs/1001.0935}{{\ttfamily arXiv:1001.0935 [nucl-th]}}.

\bibitem{Akamatsu:2011se}
Y.~Akamatsu and A.~Rothkopf, ``{Stochastic potential and quantum decoherence of
  heavy quarkonium in the quark-gluon plasma},''
  \href{http://dx.doi.org/10.1103/PhysRevD.85.105011}{{\em Phys. Rev. D}
  {\bfseries 85} (2012) 105011},
  \href{http://arxiv.org/abs/1110.1203}{{\ttfamily arXiv:1110.1203 [hep-ph]}}.

\bibitem{Gossiaux:2016htk}
P.~B. Gossiaux and R.~Katz, ``{Upsilon suppression in the
  Schrödinger--Langevin approach},''
  \href{http://dx.doi.org/10.1016/j.nuclphysa.2016.04.017}{{\em Nucl. Phys. A}
  {\bfseries 956} (2016) 737--740},
  \href{http://arxiv.org/abs/1601.01443}{{\ttfamily arXiv:1601.01443
  [hep-ph]}}.

\bibitem{Brambilla:2017zei}
N.~Brambilla, M.~A. Escobedo, J.~Soto, and A.~Vairo, ``{Heavy quarkonium
  suppression in a fireball},''
  \href{http://dx.doi.org/10.1103/PhysRevD.97.074009}{{\em Phys. Rev. D}
  {\bfseries 97} no.~7, (2018) 074009},
  \href{http://arxiv.org/abs/1711.04515}{{\ttfamily arXiv:1711.04515
  [hep-ph]}}.

\bibitem{Yao:2018nmy}
X.~Yao and T.~Mehen, ``{Quarkonium in-medium transport equation derived from
  first principles},'' \href{http://dx.doi.org/10.1103/PhysRevD.99.096028}{{\em
  Phys. Rev. D} {\bfseries 99} no.~9, (2019) 096028},
  \href{http://arxiv.org/abs/1811.07027}{{\ttfamily arXiv:1811.07027
  [hep-ph]}}.

\bibitem{Miura:2019ssi}
T.~Miura, Y.~Akamatsu, M.~Asakawa, and A.~Rothkopf, ``{Quantum Brownian motion
  of a heavy quark pair in the quark-gluon plasma},''
  \href{http://dx.doi.org/10.1103/PhysRevD.101.034011}{{\em Phys. Rev. D}
  {\bfseries 101} no.~3, (2020) 034011},
  \href{http://arxiv.org/abs/1908.06293}{{\ttfamily arXiv:1908.06293
  [nucl-th]}}.

\bibitem{Sharma:2019xum}
R.~Sharma and A.~Tiwari, ``{Quantum evolution of quarkonia with correlated and
  uncorrelated noise},''
  \href{http://dx.doi.org/10.1103/PhysRevD.101.074004}{{\em Phys. Rev. D}
  {\bfseries 101} no.~7, (2020) 074004},
  \href{http://arxiv.org/abs/1912.07036}{{\ttfamily arXiv:1912.07036
  [hep-ph]}}.

\bibitem{Vaidya:2020cyi}
V.~Vaidya and X.~Yao, ``{Transverse momentum broadening of a jet in quark-gluon
  plasma: an open quantum system EFT},''
  \href{http://dx.doi.org/10.1007/JHEP10(2020)024}{{\em JHEP} {\bfseries 10}
  (2020) 024}, \href{http://arxiv.org/abs/2004.11403}{{\ttfamily
  arXiv:2004.11403 [hep-ph]}}.

\bibitem{Yao:2020xzw}
X.~Yao, W.~Ke, Y.~Xu, S.~A. Bass, and B.~M\"uller, ``{Coupled Boltzmann
  Transport Equations of Heavy Quarks and Quarkonia in Quark-Gluon Plasma},''
  \href{http://dx.doi.org/10.1007/JHEP01(2021)046}{{\em JHEP} {\bfseries 01}
  (2021) 046}, \href{http://arxiv.org/abs/2004.06746}{{\ttfamily
  arXiv:2004.06746 [hep-ph]}}.

\bibitem{Yao:2020eqy}
X.~Yao and T.~Mehen, ``{Quarkonium Semiclassical Transport in Quark-Gluon
  Plasma: Factorization and Quantum Correction},''
  \href{http://dx.doi.org/10.1007/JHEP02(2021)062}{{\em JHEP} {\bfseries 02}
  (2021) 062}, \href{http://arxiv.org/abs/2009.02408}{{\ttfamily
  arXiv:2009.02408 [hep-ph]}}.

\bibitem{Akamatsu:2020ypb}
Y.~Akamatsu, ``{Quarkonium in quark\textendash{}gluon plasma: Open quantum
  system approaches re-examined},''
  \href{http://dx.doi.org/10.1016/j.ppnp.2021.103932}{{\em Prog. Part. Nucl.
  Phys.} {\bfseries 123} (2022) 103932},
  \href{http://arxiv.org/abs/2009.10559}{{\ttfamily arXiv:2009.10559
  [nucl-th]}}.

\bibitem{Brambilla:2020qwo}
N.~Brambilla, M.~A. Escobedo, M.~Strickland, A.~Vairo, P.~Vander~Griend, and
  J.~H. Weber, ``{Bottomonium suppression in an open quantum system using the
  quantum trajectories method},''
  \href{http://dx.doi.org/10.1007/JHEP05(2021)136}{{\em JHEP} {\bfseries 05}
  (2021) 136}, \href{http://arxiv.org/abs/2012.01240}{{\ttfamily
  arXiv:2012.01240 [hep-ph]}}.

\bibitem{Yao:2021lus}
X.~Yao, ``{Open quantum systems for quarkonia},''
  \href{http://dx.doi.org/10.1142/S0217751X21300106}{{\em Int. J. Mod. Phys. A}
  {\bfseries 36} no.~20, (2021) 2130010},
  \href{http://arxiv.org/abs/2102.01736}{{\ttfamily arXiv:2102.01736
  [hep-ph]}}.

\bibitem{Lehmann:2020fjt}
A.~Lehmann and A.~Rothkopf, ``{Proper static potential in classical lattice
  gauge theory at finite T},''
  \href{http://dx.doi.org/10.1007/JHEP07(2021)067}{{\em JHEP} {\bfseries 07}
  (2021) 067}, \href{http://arxiv.org/abs/2012.10089}{{\ttfamily
  arXiv:2012.10089 [hep-lat]}}.

\bibitem{Neill:2015nya}
D.~Neill, ``{The Edge of Jets and Subleading Non-Global Logs},''
  \href{http://arxiv.org/abs/1508.07568}{{\ttfamily arXiv:1508.07568
  [hep-ph]}}.

\bibitem{Armesto:2019mna}
N.~Armesto, F.~Dominguez, A.~Kovner, M.~Lublinsky, and V.~Skokov, ``{The Color
  Glass Condensate density matrix: Lindblad evolution, entanglement entropy and
  Wigner functional},'' \href{http://dx.doi.org/10.1007/JHEP05(2019)025}{{\em
  JHEP} {\bfseries 05} (2019) 025},
  \href{http://arxiv.org/abs/1901.08080}{{\ttfamily arXiv:1901.08080
  [hep-ph]}}.

\bibitem{Li:2020bys}
M.~Li and A.~Kovner, ``{JIMWLK Evolution, Lindblad Equation and
  Quantum-Classical Correspondence},''
  \href{http://dx.doi.org/10.1007/JHEP05(2020)036}{{\em JHEP} {\bfseries 05}
  (2020) 036}, \href{http://arxiv.org/abs/2002.02282}{{\ttfamily
  arXiv:2002.02282 [hep-ph]}}.

\bibitem{Boyanovsky:2015tba}
D.~Boyanovsky, ``{Effective field theory during inflation: Reduced density
  matrix and its quantum master equation},''
  \href{http://dx.doi.org/10.1103/PhysRevD.92.023527}{{\em Phys. Rev. D}
  {\bfseries 92} no.~2, (2015) 023527},
  \href{http://arxiv.org/abs/1506.07395}{{\ttfamily arXiv:1506.07395
  [astro-ph.CO]}}.

\bibitem{Burgess:2015ajz}
C.~P. Burgess, R.~Holman, and G.~Tasinato, ``{Open EFTs, IR effects
  \textbackslash{}\& late-time resummations: systematic corrections in
  stochastic inflation},''
  \href{http://dx.doi.org/10.1007/JHEP01(2016)153}{{\em JHEP} {\bfseries 01}
  (2016) 153}, \href{http://arxiv.org/abs/1512.00169}{{\ttfamily
  arXiv:1512.00169 [gr-qc]}}.

\bibitem{Shandera:2017qkg}
S.~Shandera, N.~Agarwal, and A.~Kamal, ``{Open quantum cosmological system},''
  \href{http://dx.doi.org/10.1103/PhysRevD.98.083535}{{\em Phys. Rev. D}
  {\bfseries 98} no.~8, (2018) 083535},
  \href{http://arxiv.org/abs/1708.00493}{{\ttfamily arXiv:1708.00493
  [hep-th]}}.

\bibitem{Zarei:2021dpb}
M.~Zarei, N.~Bartolo, D.~Bertacca, S.~Matarrese, and A.~Ricciardone,
  ``{Non-Markovian open quantum system approach to the early universe: I.
  Damping of gravitational waves by matter},''
  \href{http://arxiv.org/abs/2104.04836}{{\ttfamily arXiv:2104.04836
  [astro-ph.CO]}}.

\bibitem{Cohen:2020php}
T.~Cohen and D.~Green, ``{Soft de Sitter Effective Theory},''
  \href{http://dx.doi.org/10.1007/JHEP12(2020)041}{{\em JHEP} {\bfseries 12}
  (2020) 041}, \href{http://arxiv.org/abs/2007.03693}{{\ttfamily
  arXiv:2007.03693 [hep-th]}}.

\bibitem{Binder:2020efn}
T.~Binder, B.~Blobel, J.~Harz, and K.~Mukaida, ``{Dark matter bound-state
  formation at higher order: a non-equilibrium quantum field theory
  approach},'' \href{http://dx.doi.org/10.1007/JHEP09(2020)086}{{\em JHEP}
  {\bfseries 09} (2020) 086}, \href{http://arxiv.org/abs/2002.07145}{{\ttfamily
  arXiv:2002.07145 [hep-ph]}}.

\bibitem{cleve_et_al:LIPIcs:2017:7477}
R.~Cleve and C.~Wang, ``{Efficient Quantum Algorithms for Simulating Lindblad
  Evolution},'' in {\em 44th International Colloquium on Automata, Languages,
  and Programming (ICALP 2017)}.
\newblock \href{http://arxiv.org/abs/1612.09512}{{\ttfamily arXiv:1612.09512
  [quant-ph]}}.

\bibitem{Berges:2020fwq}
J.~Berges, M.~P. Heller, A.~Mazeliauskas, and R.~Venugopalan, ``{QCD
  thermalization: Ab initio approaches and interdisciplinary connections},''
  \href{http://dx.doi.org/10.1103/RevModPhys.93.035003}{{\em Rev. Mod. Phys.}
  {\bfseries 93} no.~3, (2021) 035003},
  \href{http://arxiv.org/abs/2005.12299}{{\ttfamily arXiv:2005.12299
  [hep-th]}}.

\bibitem{Schwinger:1962tp}
J.~S. Schwinger, ``{Gauge Invariance and Mass. 2.},''
  \href{http://dx.doi.org/10.1103/PhysRev.128.2425}{{\em Phys. Rev.} {\bfseries
  128} (1962) 2425--2429}.

\bibitem{Coleman:1975pw}
S.~R. Coleman, R.~Jackiw, and L.~Susskind, ``{Charge Shielding and Quark
  Confinement in the Massive Schwinger Model},''
  \href{http://dx.doi.org/10.1016/0003-4916(75)90212-2}{{\em Annals Phys.}
  {\bfseries 93} (1975) 267}.

\bibitem{Kharzeev:2012re}
D.~E. Kharzeev and F.~Loshaj, ``{Jet energy loss and fragmentation in heavy ion
  collisions},'' \href{http://dx.doi.org/10.1103/PhysRevD.87.077501}{{\em Phys.
  Rev. D} {\bfseries 87} no.~7, (2013) 077501},
  \href{http://arxiv.org/abs/1212.5857}{{\ttfamily arXiv:1212.5857 [hep-ph]}}.

\bibitem{Loshaj:2014aia}
F.~Loshaj and D.~E. Kharzeev, ``{Soft photon production from real-time dynamics
  of jet fragmentation},''
  \href{http://dx.doi.org/10.1016/j.nuclphysa.2014.08.006}{{\em Nucl. Phys. A}
  {\bfseries 931} (2014) 712--717},
  \href{http://arxiv.org/abs/1407.8158}{{\ttfamily arXiv:1407.8158 [hep-ph]}}.

\bibitem{Calzetta:2008iqa}
E.~A. Calzetta and B.-L.~B. Hu,
  \href{http://dx.doi.org/10.1017/CBO9780511535123}{{\em {Nonequilibrium
  Quantum Field Theory}}}.
\newblock Cambridge Monographs on Mathematical Physics. Cambridge University
  Press, 9, 2008.

\bibitem{Coleman:1976uz}
S.~R. Coleman, ``{More About the Massive Schwinger Model},''
  \href{http://dx.doi.org/10.1016/0003-4916(76)90280-3}{{\em Annals Phys.}
  {\bfseries 101} (1976) 239}.

\bibitem{Zhou:2021kdl}
Z.-Y. Zhou, G.-X. Su, J.~C. Halimeh, R.~Ott, H.~Sun, P.~Hauke, B.~Yang, Z.-S.
  Yuan, J.~Berges, and J.-W. Pan, ``{Thermalization dynamics of a gauge theory
  on a quantum simulator},''
  \href{http://dx.doi.org/10.1126/science.abl6277}{{\em Science} {\bfseries
  377} no.~6603, (2022) abl6277},
  \href{http://arxiv.org/abs/2107.13563}{{\ttfamily arXiv:2107.13563
  [cond-mat.quant-gas]}}.

\bibitem{Casher:1973uf}
A.~Casher, J.~B. Kogut, and L.~Susskind, ``{Vacuum polarization and the quark
  parton puzzle},'' \href{http://dx.doi.org/10.1103/PhysRevLett.31.792}{{\em
  Phys. Rev. Lett.} {\bfseries 31} (1973) 792--795}.

\bibitem{Banks:1975gq}
T.~Banks, L.~Susskind, and J.~B. Kogut, ``{Strong Coupling Calculations of
  Lattice Gauge Theories: (1+1)-Dimensional Exercises},''
  \href{http://dx.doi.org/10.1103/PhysRevD.13.1043}{{\em Phys. Rev. D}
  {\bfseries 13} (1976) 1043}.

\bibitem{Jordan:1928wi}
P.~Jordan and E.~P. Wigner, ``{About the Pauli exclusion principle},''
  \href{http://dx.doi.org/10.1007/BF01331938}{{\em Z. Phys.} {\bfseries 47}
  (1928) 631--651}.

\bibitem{nielsen_chuang_2010}
M.~A. Nielsen and I.~L. Chuang,
  \href{http://dx.doi.org/10.1017/CBO9780511976667}{{\em Quantum Computation
  and Quantum Information: 10th Anniversary Edition}}.
\newblock Cambridge University Press, 2010.

\bibitem{KOSSAKOWSKI1972247}
A.~Kossakowski, ``On quantum statistical mechanics of non-hamiltonian
  systems,''
  \href{http://dx.doi.org/https://doi.org/10.1016/0034-4877(72)90010-9}{{\em
  Reports on Mathematical Physics} {\bfseries 3} no.~4, (1972) 247 -- 274}.

\bibitem{Lindblad:1975ef}
G.~Lindblad, ``{On the Generators of Quantum Dynamical Semigroups},''
  \href{http://dx.doi.org/10.1007/BF01608499}{{\em Commun. Math. Phys.}
  {\bfseries 48} (1976) 119}.

\bibitem{Gorini:1976cm}
V.~Gorini, A.~Frigerio, M.~Verri, A.~Kossakowski, and E.~Sudarshan,
  ``{Properties of Quantum Markovian Master Equations},''
  \href{http://dx.doi.org/10.1016/0034-4877(78)90050-2}{{\em Rept. Math. Phys.}
  {\bfseries 13} (1978) 149}.

\bibitem{PhysRevLett.46.211}
A.~O. Caldeira and A.~J. Leggett, ``Influence of dissipation on quantum
  tunneling in macroscopic systems,''
  \href{http://dx.doi.org/10.1103/PhysRevLett.46.211}{{\em Phys. Rev. Lett.}
  {\bfseries 46} (Jan, 1981) 211--214}.

\bibitem{IBMQMontreal}
{27-qubit backend: IBM Q team, IBM Q Montreal backend specification v1.9.11,
  (2021) Retrieved from
  \href{https://quantum-computing.ibm.com}{https://quantum-computing.ibm.com}}.

\bibitem{Hu:2019}
Z.~Hu, R.~Xia, and S.~Kais, ``A quantum algorithm for evolving open quantum
  dynamics on quantum computing devices,''
  \href{http://dx.doi.org/10.1038/s41598-020-60321-x}{{\em Scientific Reports}
  {\bfseries 10} no.~1, (2020) 3301}.

\bibitem{headmarsden2020capturing}
K.~Head-Marsden, S.~Krastanov, D.~A. Mazziotti, and P.~Narang, ``Capturing
  non-markovian dynamics on near-term quantum computers,''
  \href{http://dx.doi.org/10.1103/PhysRevResearch.3.013182}{{\em Phys. Rev.
  Research} {\bfseries 3} (Feb, 2021) 013182}.

\bibitem{gupta2020optimal}
P.~Gupta and C.~Chandrashekar, ``Optimal quantum simulation of open quantum
  systems,'' \href{http://arxiv.org/abs/2012.07540}{{\ttfamily arXiv:2012.07540
  [quant-ph]}}.

\bibitem{PhysRevB.102.125112}
L.~Del~Re, B.~Rost, A.~F. Kemper, and J.~K. Freericks, ``Driven-dissipative
  quantum mechanics on a lattice: Simulating a fermionic reservoir on a quantum
  computer,'' \href{http://dx.doi.org/10.1103/PhysRevB.102.125112}{{\em Phys.
  Rev. B} {\bfseries 102} (Sep, 2020) 125112}.

\bibitem{ramusat2020quantum}
N.~Ramusat and V.~Savona, ``A quantum algorithm for the direct estimation of
  the steady state of open quantum systems,''
  \href{http://dx.doi.org/https://doi.org/10.22331/q-2021-02-22-399}{{\em
  Quantum} {\bfseries 5} (2021) 399},
  \href{http://arxiv.org/abs/2008.07133}{{\ttfamily arXiv:2008.07133
  [quant-ph]}}.

\bibitem{metcalf2021quantum}
M.~Metcalf, E.~Stone, K.~Klymko, A.~F. Kemper, M.~Sarovar, and W.~A. de~Jong,
  ``Quantum markov chain monte carlo with digital dissipative dynamics on
  quantum computers,'' \href{http://arxiv.org/abs/2103.03207}{{\ttfamily
  arXiv:2103.03207 [quant-ph]}}.

\bibitem{Qiskit}
G.~Aleksandrowicz, T.~Alexander, P.~Barkoutsos, L.~Bello, Y.~Ben-Haim,
  D.~Bucher, F.~J. Cabrera-Hernández, J.~Carballo-Franquis, A.~Chen, C.-F.
  Chen, J.~M. Chow, A.~D. Córcoles-Gonzales, A.~J. Cross, A.~Cross,
  J.~Cruz-Benito, C.~Culver, S.~D. L.~P. González, E.~D.~L. Torre, D.~Ding,
  E.~Dumitrescu, I.~Duran, P.~Eendebak, M.~Everitt, I.~F. Sertage, A.~Frisch,
  A.~Fuhrer, J.~Gambetta, B.~G. Gago, J.~Gomez-Mosquera, D.~Greenberg,
  I.~Hamamura, V.~Havlicek, J.~Hellmers, Łukasz Herok, H.~Horii, S.~Hu,
  T.~Imamichi, T.~Itoko, A.~Javadi-Abhari, N.~Kanazawa, A.~Karazeev,
  K.~Krsulich, P.~Liu, Y.~Luh, Y.~Maeng, M.~Marques, F.~J. Martín-Fernández,
  D.~T. McClure, D.~McKay, S.~Meesala, A.~Mezzacapo, N.~Moll, D.~M. Rodríguez,
  G.~Nannicini, P.~Nation, P.~Ollitrault, L.~J. O'Riordan, H.~Paik, J.~Pérez,
  A.~Phan, M.~Pistoia, V.~Prutyanov, M.~Reuter, J.~Rice, A.~R. Davila, R.~H.~P.
  Rudy, M.~Ryu, N.~Sathaye, C.~Schnabel, E.~Schoute, K.~Setia, Y.~Shi,
  A.~Silva, Y.~Siraichi, S.~Sivarajah, J.~A. Smolin, M.~Soeken, H.~Takahashi,
  I.~Tavernelli, C.~Taylor, P.~Taylour, K.~Trabing, M.~Treinish, W.~Turner,
  D.~Vogt-Lee, C.~Vuillot, J.~A. Wildstrom, J.~Wilson, E.~Winston, C.~Wood,
  S.~Wood, S.~Wörner, I.~Y. Akhalwaya, and C.~Zoufal, ``{Qiskit: An
  Open-source Framework for Quantum Computing},'' Jan., 2019.
\newblock \url{https://doi.org/10.5281/zenodo.2562111}.

\bibitem{PhysRevA.102.012426}
A.~He, B.~Nachman, W.~A. de~Jong, and C.~W. Bauer, ``Zero-noise extrapolation
  for quantum-gate error mitigation with identity insertions,''
  \href{http://dx.doi.org/10.1103/PhysRevA.102.012426}{{\em Phys. Rev. A}
  {\bfseries 102} (Jul, 2020) 012426}.

\bibitem{rattew2021quantum}
A.~G. Rattew, Y.~Sun, P.~Minssen, and M.~Pistoia, ``Quantum simulation of
  galton machines using mid-circuit measurement and reuse,''
  \href{http://arxiv.org/abs/2009.06601}{{\ttfamily arXiv:2009.06601
  [quant-ph]}}.

\bibitem{2020arXiv201000215H}
A.~{Hashim}, R.~K. {Naik}, A.~{Morvan}, J.-L. {Ville}, B.~{Mitchell}, J.~M.
  {Kreikebaum}, M.~{Davis}, E.~{Smith}, C.~{Iancu}, K.~P. {O'Brien},
  I.~{Hincks}, J.~J. {Wallman}, J.~{Emerson}, and I.~{Siddiqi}, ``{Randomized
  compiling for scalable quantum computing on a noisy superconducting quantum
  processor},'' \href{http://arxiv.org/abs/2010.00215}{{\ttfamily
  arXiv:2010.00215 [quant-ph]}}.

\bibitem{Feller2}
W.~Feller, {\em An Introduction to Probability Theory and Its Applications},
  vol.~2.
\newblock Wiley, New York, second~ed., 1971.

\bibitem{IBMQToronto}
{27-qubit backend: IBM Q team, IBM Q Toronto backend specification v1.4.24,
  (2021) Retrieved from
  \href{https://quantum-computing.ibm.com}{https://quantum-computing.ibm.com}}.

\end{thebibliography}\endgroup
\end{document}